\documentclass{article}

\usepackage[preprint]{neurips_2026}                

\usepackage[utf8]{inputenc}
\usepackage[T1]{fontenc}
\usepackage{hyperref}
\usepackage{url}
\usepackage{booktabs}
\usepackage{amsfonts}
\usepackage{amssymb}
\usepackage{nicefrac}
\usepackage{microtype}
\usepackage{xcolor}
\usepackage{graphicx}
\usepackage{xspace}
\usepackage{wrapfig}
\usepackage{boldline}
\usepackage[table]{xcolor} 
\usepackage{threeparttable}
\usepackage{graphicx}      
\usepackage{bigstrut}      
\usepackage{array}
\usepackage{xurl}
\usepackage{fvextra}
\usepackage{float}

\usepackage[most]{tcolorbox}

\newtcolorbox{errorbox}{
  colback=gray!5,
  colframe=gray!60,
  boxrule=0.5pt,
  arc=3pt,
  left=6pt,
  right=6pt,
  top=4pt,
  bottom=4pt,
  fontupper=\ttfamily\small,
  breakable
}
\usepackage{xurl} 
\newcommand{\repo}[1]{\texttt{\detokenize{#1}}}
\newcommand{\code}[1]{\texttt{\detokenize{#1}}}
\usepackage{aesthetics}

\usepackage[disable,textsize=tiny]{todonotes}
\newcommand{\scarfbench}{\textsc{ScarfBench}\xspace}
\title{\scarfbench: A Benchmark for Cross-Framework Application Migration in Enterprise Java}

\author{%
  Advait Pavuluri$^{*,1,\ddagger}$, Bridget McGinn$^{*,2}$, Ashita Saxena$^{*,2}$ \\
  \bfseries George Safta$^{2}$, Srikanth Tamilselvam$^{2}$, Raju Pavuluri$^{2, \dagger}$ \\
  \bfseries Michele Merler$^{2}$, Baishakhi Ray$^{3}$, Rahul Krishna$^{2,\dagger}$ \\
  $^{1}$Rensselaer Polytechnic Institute \quad $^{2}$IBM Software Innovation Labs \quad $^{3}$Columbia University \\
  {\normalfont\small $^{*}$Equal contribution. ~~ $^{\ddagger}$Work done at IBM. ~~ $^{\dagger}$Corresponding: \texttt{\{pavuluri@us., rkrsn@\}ibm.com}}
}

\begin{document}

\maketitle

\begin{abstract}

Java remains central to enterprise software, with many applications outliving their original architecture. Modernizing or migrating them across frameworks is required and it involves more than a local edit: it's a behavior-preserving refactoring spanning build configuration, dependency injection, persistence, request handling, and deployment. While existing software-engineering benchmarks provide strong coverage of bug fixing, feature implementation and language or version modernization, they leave cross-framework application refactoring largely unmeasured.

We introduce \scarfbench, a benchmark for behavior-preserving cross-framework
refactoring of enterprise Java applications. The benchmark is built from
expert-written implementation triples across Spring, Jakarta EE, and Quarkus:
34 applications, comprising 29 focused single-layer applications and 5 whole
applications. Together, these yield 102 framework-specific variants
($\sim$151K lines of paired Java across 1{,}946 source and test files) and 204
directed refactoring tasks. In each task, an agent receives a working source
application and a target framework and must synthesize a target implementation
that preserves the source behavior. Correctness is evaluated by an
application-specific executable oracle: the migrated candidate must compile,
deploy in a containerized target runtime and pass behavioral tests over the
application's observable interface.

We evaluate five state-of-the-art coding agents on \scarfbench{} and show that behavior-preserving framework migration remains difficult for current agents: the strongest agent achieves only $15.3\%$ aggregate test pass on focused-layer migrations and $12.2\%$ on whole applications and only one of the 204 directed migration tasks yields a fully behaviorally equivalent target. We observe that difficulty is asymmetric across framework directions and architectural layers, with Spring$\leftrightarrow$Quarkus the most tractable pair and Jakarta-targeted migrations hardest. From a combination of LLM-as-a-judge and expert adjudication of failed-task traces we derive a taxonomy of recurring failure categories spanning the build, deploy, and test stages. We release the benchmark, harness and agent traces at \url{https://scarfbench.info}.
\end{abstract}
\section{Introduction}



Java remains a persistent staple of enterprise software, with roughly 30\% of professional developers routinely shipping code in Java~\cite{stackoverflow-survey-2024,newrelic-java-2024}. Most enterprise Java applications are authored in one of three widely used frameworks, namely Spring, Jakarta EE and Quarkus, which together account for a large share of JVM application-framework usage in practice~\cite{snyk-jvm-2021,jetbrains-devstate-2024}. These systems are long-lived, remaining in production well beyond the obsolescence of the architectural assumptions that guided their original implementation. This forces periodic, risk-prone migrations to more modern stacks, driven both by deprecation pressure and by non-functional goals such as lower memory overhead, faster startup and improved cloud elasticity~\cite{logicmonitor-quarkus}.

\begin{figure}[t!]
\centering
  \includegraphics[width=0.9\linewidth]{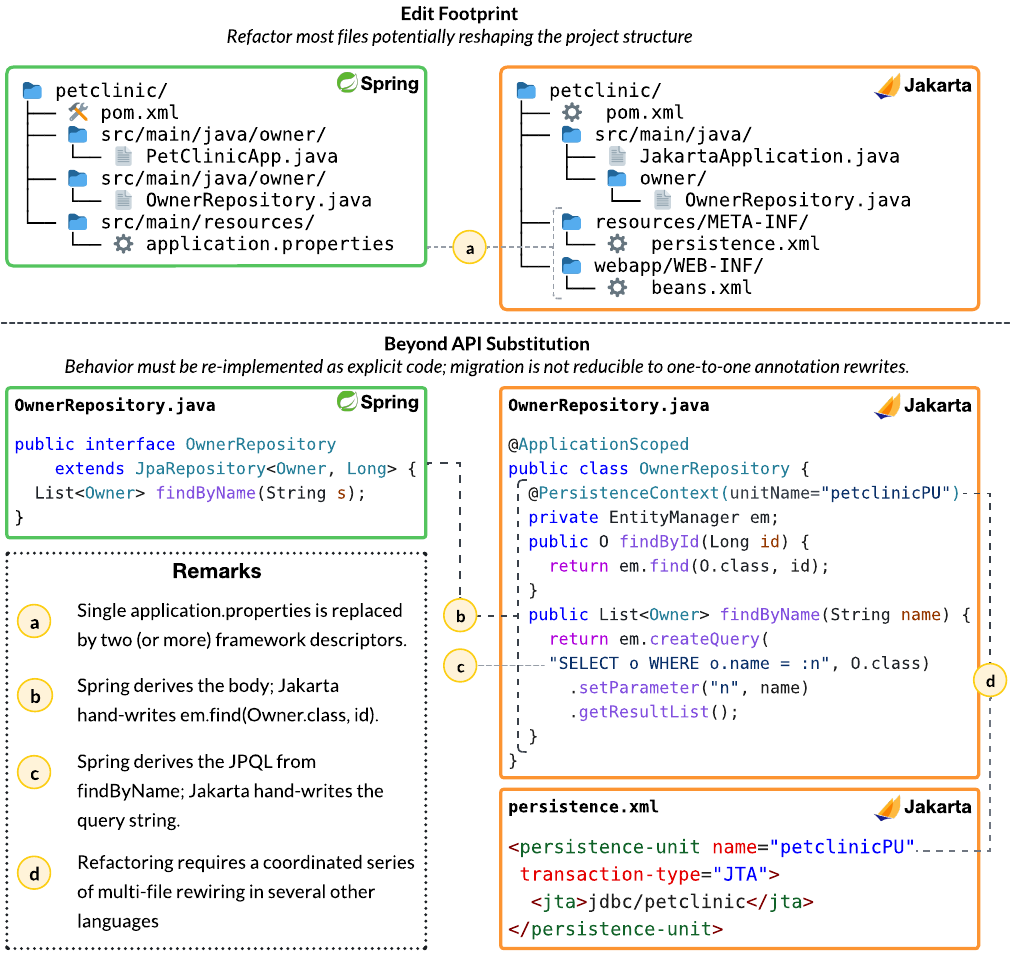}
  \caption{Migration is a \textit{structural transformation} across heterogeneous artifacts: porting Spring to Jakarta expands a 3-line interface into a 14-line CDI bean, rewrites derived queries as
  hand-written JPQL, externalizes auto-config into JPA and CDI descriptors and adds a Java–XML string binding.}
  \label{fig:anatomy-intro}
\end{figure}

However, framework migration is difficult because it requires coordinated changes across many interconnected parts of an application. Developers must not only rewrite framework-specific APIs, but also preserve the application’s behavior while adapting build configurations, dependency injection, persistence, request handling, security, and deployment to a new framework’s programming model and runtime assumptions~\cite{logicmonitor-quarkus}. Figure~\ref{fig:anatomy-intro} illustrates this on a single data-access class: porting it from Spring to Jakarta replaces a thin repository interface with a hand-written CDI bean, externalizes auto-configured behavior into separate descriptor files, and introduces unchecked Java--XML name bindings. These changes are tightly coupled: even a small mistake in one layer can break compilation, prevent deployment, or introduce subtle runtime errors that are difficult to detect~\cite{keycloak-migration,vodafone-greece,orange-quarkus,carrefour-quarkus,lufthansa-aviatar,stargate}.

Despite recent progress in software-engineering benchmarks to evaluate agentic capabilities of LLMs, this kind of architectural migration remains largely unmeasured. Existing coding-agent benchmarks primarily evaluate issue resolution, code completion, or feature implementation within a fixed language and framework stack~\cite{swebench,multi-swe-bench,repobench,fea-bench}. Migration-oriented benchmarks move closer to our setting, but mainly study version upgrades or dependency modernization, such as migrating Java~8 applications to newer LTS releases~\cite{liu2025migrationbenchrepositorylevelcodemigration,freshbrew,jmigbench} or repository modernization through implementation-agnostic testing~\cite{repomodbench}. These tasks typically preserve the same underlying framework abstractions and runtime model, an assumption that does not hold under cross-framework migration.


In this work, we present ScarfBench (\underline{S}elf-\underline{C}ontained \underline{A}pplication \underline{R}e\underline{f}actoring \underline{Bench}mark), a benchmark for behavior-preserving framework migration by LLM agents. Each task gives an agent a working application in one framework and asks it to produce an equivalent implementation in another, across Spring, Jakarta EE, and Quarkus. For each application family, expert Java developers implemented all three framework variants, yielding a runnable source and a validated human-written target for every directed migration. ScarfBench comprises 34 application families and 102 framework-specific variants. Its \textit{focused} tier contains 29 self-contained applications spanning five JSR-anchored layers of the enterprise Java stack: dependency injection, persistence, presentation, infrastructure, and business-domain logic~\citep{jsr365,jsr338,jsr356,jsr369,jsr370,jsr236,jsr345}. Its \textit{whole-application} tier adds five larger open-source systems that combine these layers in realistic use cases. This yields 204 directed migration tasks across the six framework pairs. Each candidate is rebuilt and tested in a containerized harness against the target runtime and 1{,}331 expert-written tests. A task passes only if it compiles, deploys, and preserves behavior.

We evaluate five state-of-the-art coding agents powered by frontier models on ScarfBench: Claude Code with Claude Opus 4.6, Gemini CLI with Gemini-3.1 Pro, Codex with GPT-5.4, Opencode with GLM-5.1, and Qwen CLI with Qwen3.5-397B-A17B. Our evaluation
reveals three main findings.

\noindent\textbf{Finding 1: Cross-framework migration is beyond current agents.}
Across the 204 directed migration tasks, the strongest agent achieves
only $15.3\%$ aggregate test pass on focused-layer migrations and
$12.2\%$ on whole applications, and only one agent-produced migration
passes its full test suite.

\noindent\textbf{Finding 2: Building and deploying do not predict correctness.}
Agents frequently produce target-framework code that compiles and
deploys yet fails the original test suite, indicating that surface-level
translation success is not sufficient for behavioral preservation and
that build- or deploy-only oracles substantially overstate migration
quality.

\noindent\textbf{Finding 3: Difficulty is asymmetric across migration targets.}
  Pass rates vary sharply by target framework: only $2\%$ of migrations to
  Jakarta EE pass behavioral tests, against $12\%$ for Spring and $14\%$
  for Quarkus, with $57\%$ of Jakarta-targeted attempts failing already
  at the compile gate.

\noindent\textbf{Contributions.} This work makes the following contributions:
\begin{itemize}
\item \textbf{ScarfBench}, an expert-validated benchmark for cross-framework
migration in enterprise Java: 34 application families across Spring,
Jakarta EE, and Quarkus yielding 102 variants ($\sim$151K lines of
paired Java) and 204 directed migration tasks, each requiring
edits to a median of 11 files and 370 added/removed lines (and up to
$>$200 files and $>$14{,}000 lines on the whole-application tier),
scored by 1{,}331 expert-written tests in a 
containerized harness with strict compile/deploy/test-pass criteria.
\item \textbf{An empirical evaluation of five state-of-the-art coding
agents} on ScarfBench, characterizing per-direction and per-JSR-layer
difficulty asymmetries.
\item \textbf{A failure-mode taxonomy} of 13 categories spanning the
build, deploy, and test stages, induced by expert developers from
failed-task traces across 5 agents $\times$ 204 directed tasks and
applied at scale by independent LLM annotators (inter-annotator
Cohen's $\kappa = 0.72$, with disagreements resolved by expert
adjudication), characterizing how cross-framework migration breaks
down for current agents.
\end{itemize}

\section{Related Work}

\paragraph{Software engineering benchmarks.}Despite the growth of software-engineering benchmarks \citep{jiang2025agenticsoftwareissueresolution,li2026advancesfrontiersllmbasedissue,guo2025comprehensivesurveybenchmarkssolutions}, only a small subset is routinely used to evaluate agentic coding systems \citep{jain2024livecodebenchholisticcontaminationfree, terminalbench}. Existing benchmarks focus primarily on Python issue resolution \citep{deng2025swebenchproaiagents,jimenez2024swebench}, with limited Java and enterprise application coverage. Representative Java-oriented benchmarks include SWE-rebench V2 \citep{badertdinov2026swerebenchv2languageagnosticswe}, Multi-SWE-bench \citep{zan2025multiswebenchmultilingualbenchmarkissue}, SWE-bench-java \citep{zan2024swebenchjavagithubissueresolving}, SWE-Bench Multilingual \citep{swesmith}, SWE-PolyBench \citep{rashid2025swepolybenchmultilanguagebenchmarkrepository}, OmniGIRL \citep{guo2025omnigirlmultilingualmultimodalbenchmark}, and OmniCode \citep{sonwane2026omnicodebenchmarkevaluatingsoftware}. Table~1 summarizes the capability coverage of these and other Java specific benchmarks.

Most existing datasets are dominated by standalone libraries and developer tools rather than full-stack enterprise applications. Cloud-native frameworks, enterprise middleware, deployment environments, and cross-framework migration scenarios remain sparsely represented.

\begin{wraptable}{r}{0.5\linewidth}
\centering
\scriptsize
\setlength{\tabcolsep}{4pt}
\renewcommand{\arraystretch}{1.15}
\caption{Capability-level comparison with representative software-engineering and migration benchmarks. A checkmark indicates that the benchmark directly targets the capability; a tilde indicates partial or incidental coverage.}
\label{tab:benchmark-capability-matrix}

\resizebox{\linewidth}{!}{%
\begin{tabular}{lccccccc}
\toprule
\rotatebox[origin=lB]{0}{\textbf{Benchmark}} &
\rotatebox[origin=lB]{90}{\textbf{Java}} &
\rotatebox[origin=lB]{90}{\textbf{App-level}} &
\rotatebox[origin=lB]{90}{\textbf{Enterprise}} &
\rotatebox[origin=lB]{90}{\textbf{Migration}} &
\rotatebox[origin=lB]{90}{\textbf{Cross-framework}} &
\rotatebox[origin=lB]{90}{\textbf{Deploy}} &
\rotatebox[origin=lB]{90}{\textbf{Behavior tests}} \\
\midrule
SWE-bench (Verified) & $\cdot$ & \checkmark & $\cdot$ & $\cdot$ & $\cdot$ & $\sim$ & $\sim$ \\
SWE-rebench V2 & \checkmark & \checkmark & $\sim$ & $\cdot$ & $\cdot$ & $\sim$ & $\sim$ \\
SWE-bench-java & \checkmark & \checkmark & $\sim$ & $\cdot$ & $\cdot$ & $\sim$ & $\sim$ \\
Multi-SWE-bench & \checkmark & \checkmark & $\cdot$ & $\cdot$ & $\cdot$ & $\sim$ & $\sim$ \\
SWE-PolyBench & \checkmark & \checkmark & $\sim$ & $\cdot$ & $\cdot$ & $\sim$ & $\sim$ \\
JavaBench & \checkmark & $\sim$ & $\cdot$ & $\cdot$ & $\cdot$ & $\cdot$ & $\sim$ \\
CoReQA & $\sim$ & \checkmark & $\sim$ & $\cdot$ & $\cdot$ & $\cdot$ & $\cdot$ \\
LongContextQA & $\sim$ & \checkmark & $\sim$ & $\cdot$ & $\cdot$ & $\cdot$ & $\cdot$ \\
MigrationBench & \checkmark & \checkmark & $\cdot$ & \checkmark & $\cdot$ & $\sim$ & $\sim$ \\
JMigBench & \checkmark & $\cdot$ & $\cdot$ & \checkmark & $\cdot$ & $\cdot$ & $\cdot$ \\
FreshBrew & \checkmark & \checkmark & $\sim$ & \checkmark & $\cdot$ & $\sim$ & $\sim$ \\
JavaBackports & \checkmark & $\sim$ & $\cdot$ & \checkmark & $\cdot$ & $\cdot$ & $\cdot$ \\
BeyondSWE & \checkmark & \checkmark & $\cdot$ & \checkmark & $\cdot$ & $\sim$ & $\sim$ \\
RepoMod-Bench & \checkmark & \checkmark & $\cdot$ & \checkmark & $\cdot$ & \checkmark & \checkmark \\
Spring AI Agent Bench & \checkmark & \checkmark & $\sim$ & $\cdot$ & $\cdot$ & $\sim$ & $\sim$ \\
\textbf{\scarfbench} & \checkmark & \checkmark & \checkmark & \checkmark & \checkmark & \checkmark & \checkmark \\
\bottomrule
\end{tabular}%
}
\vspace{0.5em}
\footnotesize
\checkmark = targeted; $\sim$ = partial/incidental; $\cdot$ = not a primary focus.
\end{wraptable}

Beyond issue resolution, Java-focused benchmarks remain limited in scope. Existing work targets object-oriented code generation~\cite{cao2024javabenchbenchmarkobjectorientedcode}, class-level test generation~\cite{zhang2024testbenchevaluatingclassleveltest,Lops_2025}, and code understanding~\cite{dhulshette2025hierarchicalrepositorylevelcodesummarization,chen2025coreqauncoveringpotentialslanguage,maharaj2026robustnessreasoningfidelitylarge}, with little focus on enterprise-specific concerns. Spring AI Agent Bench~\cite{springai2026agentbench} is the first benchmark built within an enterprise application framework, but it is limited to Spring tooling. Overall, existing benchmarks provide limited coverage of framework-specific reasoning and the multi-module, configuration-heavy systems common in enterprise Java applications.

\noindent\textbf{Migration / modernization benchmarks.}
Existing migration benchmarks primarily focus on version upgrades \cite{misra2025gitchameleon20evaluatingai, magesty2026promiseawait}, library migrations \cite{barbosa2026mig4}, and dependency-driven updates \cite{beyondswe2026}. Java-specific datasets such as MigrationBench \cite{liu2025migrationbenchrepositorylevelcodemigration}, JMigBench \cite{jmigbench}, and FreshBrew \cite{freshbrew} mainly target Java version upgrades and environment adaptation tasks \cite{cheng2025codemenvbenchmarkinglargelanguage}, while Java backporting has also been studied \cite{kahapola2026javabackports, zhong2025backportbenchmultilingualbenchmarkautomated}. Although non-trivial, these tasks largely preserve the underlying programming model and framework structure.

AI-assisted modernization has been studied from reliability and security perspectives \cite{ponnusamy2025applicationmodernizationllmsaddressing}, but without a reproducible benchmark. In contrast, our work addresses \emph{cross-framework} enterprise Java migration, coordinating changes to dependency injection, configuration, persistence, security, reactive stacks, and build toolchains while preserving application behavior.


\section{\scarfbench}
%
%
%

\scarfbench conceptualizes enterprise Java framework migration as behavior-preserving application refactoring. Each benchmark task involves migrating a self-contained application, initially implemented with Spring, Quarkus, or Jakarta EE, to a designated target framework. For every task, expert developers provide behaviorally equivalent implementations across all three frameworks. A hidden containerized evaluation harness automatically rebuilds submitted solutions and executes behavioral test suites derived from plain-language specifications.

\subsection{Task Formulation}

The basic unit of \scarfbench is a directed refactoring task. The refactoring task start from a source application written in one of three frameworks (Spring, Jakarta EE, or Quarkus) and requires that the same application behavior is re-expressed in the target framework using idiomatic patterns native to the target. Framework migration is a non-symmetric transformation in that the task $f_s \rightarrow f_t$ of migrating from source $f_s$ to target $f_t$ is distinct from the task $f_t \rightarrow f_s$ of migrating in the opposite direction. For example, Spring's inversion-of-control container and annotation-based configuration style are quite different from Jakarta EE's convention-over-configuration approach, so a migration from Spring to Jakarta EE is not simply the reverse of a migration from Jakarta EE to Spring.

Let $\mathcal{F}=\{\textsc{Spring}, \textsc{Jakarta EE}, \textsc{Quarkus}\}$ be the set of supported frameworks. Let $\mathcal{A}$ denote the set of application families in \scarfbench. For each application family $a\in\mathcal{A}$ and framework $f\in\mathcal{F}$, \scarfbench contains an implementation $I_{a,f}$ of $a$ in $f$. A task instance is an ordered triple: $\tau=(a,f_s,f_t)$;
where $a\in\mathcal{A}$, $f_s,f_t\in\mathcal{F}$, and $f_s\neq f_t$, where the input is the source implementation $I_{a,f_s}$ and the target framework $f_t$. The required output is a migrated implementation $\hat I_{a,f_t}$ that preserves the behavior of $a$ while using the target's native programming model and idioms.

Each application family $a \in \mathcal{A}$ has a shared behavioral oracle
$\mathcal{O}_a$ across all framework implementations. The oracle consists of
developer-written BDD-style test cases and concrete Playwright\footnote{\url{https://playwright.dev/}}
scripts that check externally observable equivalence. During evaluation, the
migrated candidate $\hat I_{a,f_t}$ is rebuilt and deployed in the target
container, and the oracle produces
$\mathcal{O}_a(\hat I_{a,f_t})=(O_c,O_d,O_t)$, where $O_c,O_d\in\{0,1\}$
denote build and startup success, and $O_t\in[0,1]$ is the fraction of
behavioral tests that pass. These signals are gated: $O_d=0$ if $O_c=0$, and
$O_t=0$ if $O_d=0$. Task success requires strict behavioral equivalence,
measured as $\mathbf{1}[O_t=1]$. Section \ref{sec:evaluation-protocol} describes how these
signals are aggregated across the corpus.

\subsection{Benchmark Curation and Construction}

\begin{figure}[t!]
\centering
\includegraphics[width=0.95\linewidth]{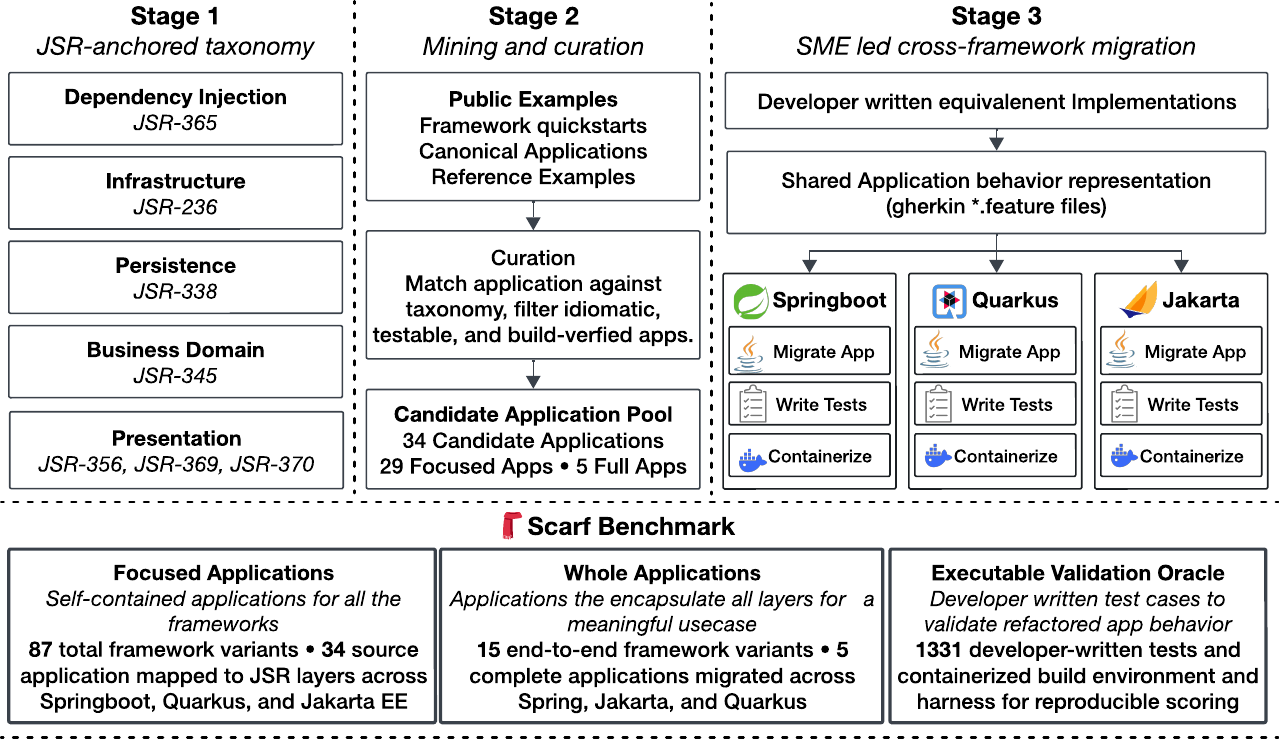}
\caption{Overview of ScarfBench construction: 34 application families implemented across Spring, Jakarta EE and Quarkus producing 102 variants accompanied by 1,331 expert-written tests in a reproducible containerized harness with strict compile/deploy/test-pass criteria. 
}
\label{fig:scarf-overview}
\end{figure}

\scarfbench{} is a paired corpus of application families implemented across Spring Boot, Quarkus, and Jakarta EE (Figure~\ref{fig:scarf-overview}). It has two tiers: \textit{focused applications} isolate individual framework concerns, while \textit{whole applications} expose the cross-layer coupling that makes migration more than local API rewrites. Applications are drawn from official framework examples, such as Eclipse's Jakarta EE examples, and maintainer-canonical repositories, such as \texttt{spring-projects/spring-petclinic}. We exclude examples that are primarily framework-neutral Java, lack externally testable behavior, or require manual setup beyond containerization and platforming. Appendix~\ref{app:dataset} details all 34 application families, their behavioral test suites, and the scale of the resulting 204 directed migration tasks.

\textbf{Focused tier.}
This tier contains 29 self-contained applications, each chosen to isolate a standard layer of the enterprise Java stack. The applications come from the Eclipse Foundation's official Jakarta EE examples repository,\footnote{\url{https://github.com/eclipse-ee4j/jakartaee-examples}} which provides one-technology demonstrations for the Jakarta EE Tutorial. We anchor this tier in Java Specification Requests (JSRs), the Java Community Process specifications that defined enterprise Java APIs. The selected layers cover migration-relevant concerns including dependency injection, managed concurrency, persistence, business services, and presentation-layer HTTP, REST, and WebSocket~\citep{jsr365}.

\textbf{Whole-application tier.}
This tier contains five larger applications that combine these layers into end-to-end enterprise use cases. They capture interactions that isolated examples miss, such as persistent state flowing through REST endpoints or templates, container-managed services coordinating with transactions, and external services such as Kafka, JMS, or PostgreSQL expanding the deployment and configuration surface. These tasks test whether agents can re-express application behavior through the target framework's runtime model and idioms, rather than only translate framework APIs.

\textbf{Paired implementation workflow.}
For each accepted application, enterprise Java experts manually implemented variants in all three frameworks, preserving behavior while using each target framework's native idioms. Each variant was compiled, deployed, containerized, and validated against the behavioral oracle before inclusion. Figure~\ref{fig:anatomy-intro} illustrates one such expert-authored migration: PetClinic from Spring Boot to Jakarta. The migration spans build files, bootstrap code, routing, dependency injection, repositories, configuration, and templates, while preserving cross-file invariants such as route-template links and injected configuration keys. These are the behavior whole-application tasks are designed to expose.

Appendix~\ref{appendix:daytrader-spring-quarkus} provides a detailed case study of the DayTrader Spring-to-Quarkus migration workflow, including dependency mapping, configuration translation, code-transformation patterns, and validation procedures.

\subsection{Behavioral test suites}
\label{sec:behavioral-test-suites}
Each application $a \in \mathcal{A}$ is paired with a framework-neutral behavioral oracle $\mathcal{O}_a$ that defines equivalence at the externally observable boundary: HTTP routes, response payloads, UI flows, validation outcomes, and persistent state changes. We use this boundary rather than unit tests or source-level checks because framework migration intentionally changes class layout, dependency-injection fixtures, test runners, and package structure; reusing source-framework unit tests would require migrating the oracle itself.

Each oracle is an atomic behavioral scenario with an initial state, a user- or protocol-level operation, and an expected observable outcome. Scenarios are first defined once per application family as a Gherkin feature file, then concretized as Playwright/pytest tests for each framework variant. These concretizations run the same scenarios against containers and remain nearly identical across Spring, Jakarta EE, and Quarkus, differing only when a framework exposes a genuinely different external convention. Such visible conventions are part of the contract; for example, JSF pages served under a \texttt{.xhtml} path must retain that path after migration. Appendix~\ref{appendix:daytrader-oracle} illustrates the oracle construction process for DayTrader by mapping framework-independent Gherkin specifications to executable Spring and Quarkus smoke tests.

The migration authors also wrote the behavioral scenarios and executable concretizations during benchmark construction. An application enters \scarfbench{} only after all three expert-authored framework variants compile, deploy, and pass the complete oracle in the containerized harness. Thus, each oracle must accept three idiomatically distinct implementations of the same externally observable behavior before it is used to evaluate agent-generated migrations.

\section{Evaluation Protocol}
\label{sec:evaluation-protocol}

\textbf{Inference setup.}~
We evaluate five coding-agent stacks using their standard repository-editing, shell, build, and test tools. Each agent receives the source variant $I_{a,f_s}$ and target framework specification, and produces a candidate migration $\hat I_{a,f_t}$; agents do not receive the expert-written target implementation $I_{a,f_t}$. Following SWE-smith~\citep{swesmith}, each (agent, task) pair is run once at temperature~0; we report pass@1 and perform no inference-time scaling. We evaluate Claude Code with Claude Opus~4.6, Codex with GPT-5.4, Gemini CLI with Gemini-3.1 Pro, Qwen CLI with Qwen-3.5-397B, and OpenCode with GLM-5.1. Evaluation costs are reported in Appendix~\ref{app:evaluation-cost}.

\textbf{Prompt-packaging variants.}~
For each agent stack, we compare two prompt variants: a monolithic prompt containing all instructions, and a skills-based directory that modularizes guidance and tool specifications. This lets us measure the effect of prompt organization on migration performance; prompts are described in Appendix~\ref{app:prompts}.

\textbf{Execution harness.}~
Each candidate is packaged into the target framework's containerized runtime using a framework-specific Maven build and Docker configuration. Focused tasks run as a single application container; whole-application tasks that require external services use Docker Compose to launch the application with its dependencies, such as databases or message brokers. This standardizes the build toolchain and test driver across agents and tasks.

For a task $\tau=(a,f_s,f_t)$, the harness evaluates the migrated candidate $\hat I_{a,f_t}$ against the framework-neutral oracle $\mathcal{O}_a$ in three sequential stages:
\begin{enumerate}
    \item \textbf{Compile.} The target image must build successfully from the candidate sources and declared dependencies. We define $C_{\tau}\in\{0,1\}$, where $C_{\tau}=1$ iff the build succeeds.

    \item \textbf{Deploy.} If compilation succeeds, the application container or Compose stack must start within timeout and emit the expected readiness signal. We define $D_{\tau}\in\{0,1\}$, where $D_{\tau}=1$ iff startup succeeds.

    \item \textbf{Behavioral tests.} If deployment succeeds, the harness runs the oracle tests for application $a$. Let $A_a$ be the set of behavioral assertions in $\mathcal{O}_a$, and let $B_{\tau,j}\in\{0,1\}$ indicate whether assertion $j\in A_a$ passes for $\hat I_{a,f_t}$. The assertion pass rate is
    $S_{\tau}=|A_a|^{-1}\sum_{j\in A_a} B_{\tau,j}$; if compilation or deployment fails, $S_{\tau}=0$.
\end{enumerate}

We record the first failing stage for error diagnosis. Test pass rate is the primary behavioral correctness signal. It jointly reflects build correctness, deployability, and preservation of externally observable behavior. 

\section{Experimental Results}
\begin{table*}
\caption{\textbf{Aggregate SCARF leaderboard.} Pass@1 rates over all six
directed framework migrations for whole applications and focused apps. Columns
report compile (\textit{c}), run/deploy (\textit{r}) and test (\textit{t})
success. \colorbox{blue!18}{Highlighted} cells mark the per-column maximum
within each setting. Claude Code with Opus-4.6 is the strongest whole-app run,
reaching $87\%$ compile, $40\%$ deploy and $12\%$ test success; on focused apps
it leads on compile ($93\%$) while Gemini CLI leads on deploy and test
($61\%$, $15\%$); skills help Gemini most ($7\%\!\to\!61\%$ deploy). Aggregate progression visualizations for compile, deploy, and behavioral success rates are provided in Appendix~\ref{app:aggregate-visualization}.}
\label{tab:scarf-leaderboard-aggregate}
\centering
\scriptsize
\setlength{\tabcolsep}{4pt}
\resizebox{0.56\linewidth}{!}{%
\begin{tabular}{llrrrV{2}rrr}
\clineB{3-8}{2}
\multicolumn{2}{c}{} &
\multicolumn{6}{c}{\textsc{\textbf{with skills}}} \bigstrut\\\clineB{3-8}{2}
\multicolumn{2}{c}{} &
	\multicolumn{3}{cV{2}}{\textit{whole}} &
	\multicolumn{3}{c}{\textit{focused}} \bigstrut\\\clineB{3-8}{2}
Harness & Model &
{c~} & {r~} & {t~} &
{c~} & {r~} & {t~} \bigstrut\\\clineB{3-8}{2}
       &              & \multicolumn{1}{c}{} & \multicolumn{1}{c}{} & \multicolumn{1}{c}{} & \multicolumn{1}{c}{} & \multicolumn{1}{c}{} & \multicolumn{1}{c}{} \bigstrut\\[-1.4em]\hlineB{2}
    Claude Code & Opus-4.6  & \cellcolor{blue!18}{\bfseries{\color{blue!18}0}87} & \cellcolor{blue!18}{\bfseries{\color{blue!18}0}40} & \cellcolor{blue!18}{\bfseries{\color{blue!18}0}12} & \cellcolor{blue!18}{\bfseries{\color{blue!18}0}93} & 28          & 7 \bigstrut\\
	Gemini CLI  & G-3.1-Pro & 53 & 7 & 0 & 87          & \cellcolor{blue!18}{\bfseries{\color{blue!18}0}61} & \cellcolor{blue!18}{\bfseries{\color{blue!18}0}15} \bigstrut\\
	Codex       & GPT-5.4   & 27 & 10 & 2 & 79          & 52          & 14 \bigstrut\\
	Opencode    & GLM-5.1   & 57 & 7 & 0 & 72          & 52          & 14 \bigstrut\\
	Qwen CLI    & QW3.5-397B & 50 & 10 & 3 & 74          & 36          & 11 \bigstrut\\\hlineB{2}
\end{tabular}%
}
\resizebox{0.304\linewidth}{!}{%
\begin{tabular}{rrrV{2}rrr}
\clineB{1-6}{2}
\multicolumn{6}{c}{\textsc{\textbf{no skills}}} \bigstrut\\\clineB{1-6}{2}
	\multicolumn{3}{cV{2}}{\textit{whole}} &
	\multicolumn{3}{c}{\textit{focused}} \bigstrut\\\clineB{1-6}{2}
{c~} & {r~} & {t~} &
{c~} & {r~} & {t~} \bigstrut\\\clineB{1-6}{2}
\multicolumn{1}{c}{} & \multicolumn{1}{c}{} & \multicolumn{1}{c}{} &
\multicolumn{1}{c}{} & \multicolumn{1}{c}{} & \multicolumn{1}{c}{} \bigstrut\\[-1.4em]\hlineB{2}
        \cellcolor{blue!18}{\bfseries{\color{blue!18}0}57} & \cellcolor{blue!18}{\bfseries{\color{blue!18}0}30} & \cellcolor{blue!18}{\bfseries{\color{blue!18}0}5} & \cellcolor{blue!18}{\bfseries{\color{blue!18}0}47} & \cellcolor{blue!18}{\bfseries{\color{blue!18}0}24} & \cellcolor{blue!18}{\bfseries{\color{blue!18}0}12} \bigstrut\\
		47 & 13 & 2 & 17 & 7 & 2 \bigstrut\\
		40 & 17 & 2 & 27 & 10 & 3 \bigstrut\\
		33 & 0 & 0 & 21 & 10 & 6 \bigstrut\\
		13 & 0 & 0 & 22 & 13 & 5 \bigstrut\\\hlineB{2}
\end{tabular}%
}
\end{table*}

\subsection{How well do agents perform on cross-framework migration?}

We evaluate five state-of-the-art coding agents on the 204 directed migration tasks in \scarfbench{}, reporting compile, deploy, and test pass rates for whole- and focused-application settings, with and without skills. Across agents and settings, many candidates compile, fewer deploy, and only a small fraction pass the behavioral test suite.

\begin{wrapfigure}{r}{0.6\linewidth}
    \centering
    \includegraphics[width=\linewidth]{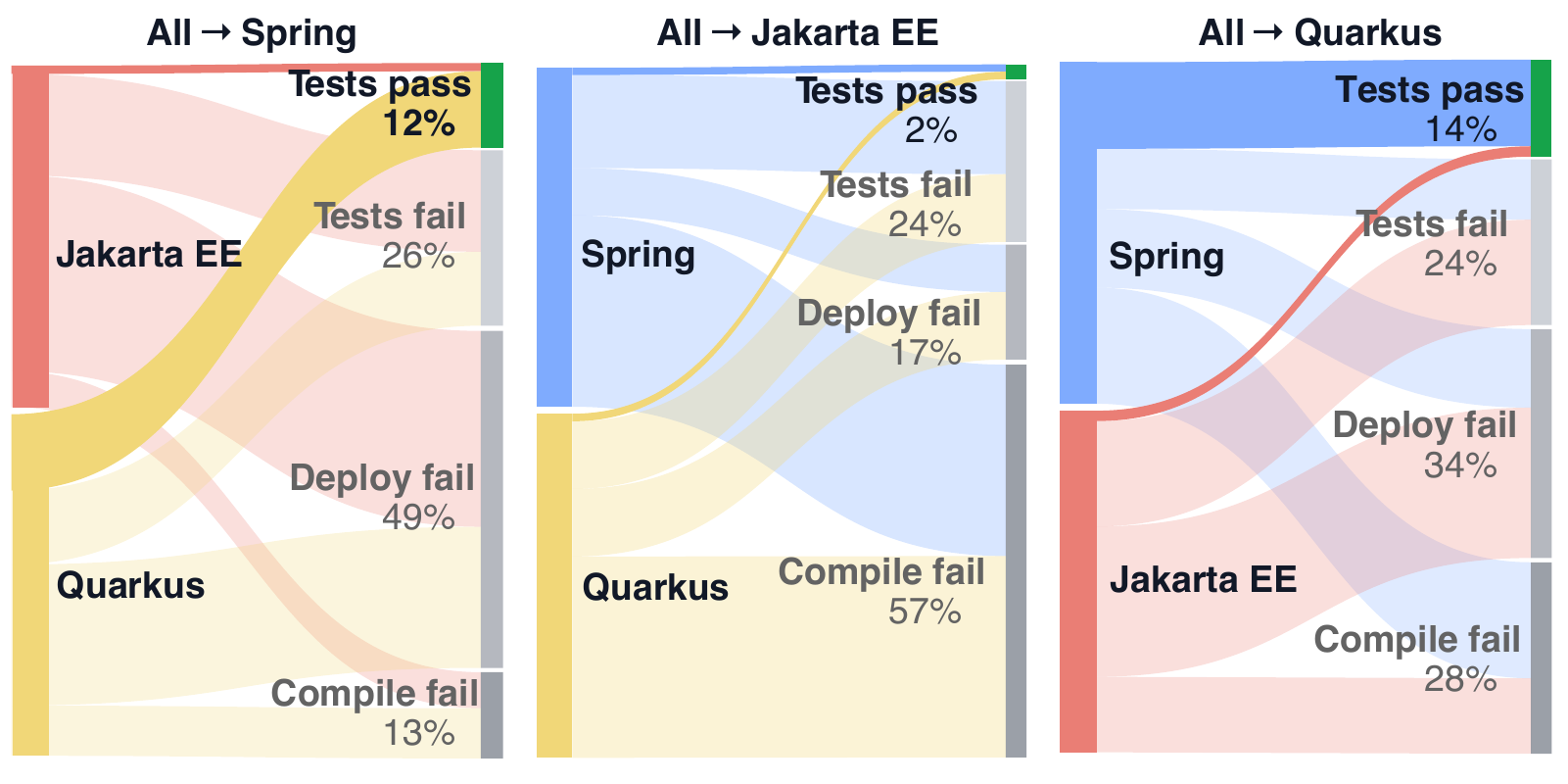}
    \caption{Each panel routes source-framework migration attempts. Jakarta EE is the hardest target: only $2\%$ of migrations from Spring or Quarkus pass behavioral tests, against $12\%$ for Spring and $14\%$ for Quarkus.}
    \label{tab:sankey}
    \vspace{-1em}
\end{wrapfigure}

Table~\ref{tab:scarf-leaderboard-aggregate} reports aggregate agent-level results. In Figure~\ref{tab:sankey}, we summarize the migration success by target framework. We offer detailed per-direction breakdowns in Appendix~\ref{app:leaderboard-breakdown},
Tables~\ref{tab:scarf-leaderboard-skills}
and~\ref{tab:scarf-leaderboard-no-skills}. Our experiments reveal several key insights about the state of cross-framework migration:

 \textbf{Whole-application migrations remain difficult.} Whole applications require
    coordinated changes across configuration, dependency wiring, persistence, routing, and
    framework runtime behavior. Even the strongest whole-application reaches only
    $12\%$ test success, while most agents remain near zero.

 \textbf{Focused migrations are more tractable but still far from solved.} Agents achieve
    higher compile and deploy rates on focused applications, but behavioral success remains limited,
    with the best focused setting reaching $15\%$ test success. Isolating a migration layer reduces
    coordination burden, but does not eliminate semantic drift or incomplete framework adaptation.

 \textbf{Skills improve compilability more than behavioral correctness.} Skills-based prompting improves compile and deploy outcomes, especially in focused settings, but the gains shrink at the test stage. Skills help agents produce more runnable migrations but not working ones.

\textbf{Migration difficulty is target-dependent and asymmetric.} Per-direction results
show that Spring$\leftrightarrow$Quarkus migrations are generally more tractable,
while Jakarta-targeted migrations are the hardest: only $2\%$ of migrations to
Jakarta EE pass behavioral tests, below Spring ($12\%$) and Quarkus ($14\%$);
$57\%$ fail at compile, versus $13\%$ and $28\%$ for the other targets
(Figure~\ref{tab:sankey}). Failures concentrate at the build gate, suggesting
that Jakarta's packaging and server-configuration conventions diverge most
sharply from learned framework idioms.

\subsection{Why do agents fail?}
\label{sec:failure-taxonomy}
\begin{table*}[!tbp]
\caption{\textbf{Per-agent failure-mode distribution within each phase.} Each (agent, slice (\textit{w} or \textit{f}), phase) block of cells sums to ${\sim}100\%$. \textit{w}: whole-app
  migrations (30/agent); \textit{f}: focused (174/agent). Transient and unrelated errors are excluded. Detailed failure taxonomy breakdowns and representative error traces are provided in Appendix~\ref{app:taxonomy-deepdive}.}
\label{tab:taxonomy-heatmap}
\centering
\scriptsize
\setlength{\tabcolsep}{3pt}

\newcommand{\hmbox}[3]{%
  \cellcolor{#1}\makebox[1.7em][c]{\strut\textcolor{#2}{#3}}%
}

\newcommand{\hm}[1]{%
  \ifnum#1=0
    \hmbox{gray!8}{gray!60}{0}%
  \else
    \pgfmathtruncatemacro{\shade}{min(100,max(1,round(100*(#1/100)^0.99)))}%
    \edef\hmfill{red!\shade}%
    \ifnum\shade<48
      \expandafter\hmbox\expandafter{\hmfill}{black}{#1}%
    \else
      \expandafter\hmbox\expandafter{\hmfill}{white}{#1}%
    \fi
  \fi
}

\newcommand{\hmnd}{\hmbox{gray!8}{gray!60}{$\cdot$}}

\resizebox{0.85\linewidth}{!}{%
\begin{tabular}{@{}>{\centering\arraybackslash}m{0.6cm} l V{2} rr V{2} rr V{2} rr V{2} rr V{2} rr@{}}
\clineB{3-12}{2}
\multicolumn{2}{c}{} & \multicolumn{2}{cV{2}}{\texttt{Claude Code}}
  & \multicolumn{2}{cV{2}}{\texttt{Gemini CLI}}
  & \multicolumn{2}{cV{2}}{\texttt{Codex}}
  & \multicolumn{2}{cV{2}}{\texttt{OC-GLM5.1}}
  & \multicolumn{2}{c}{\texttt{Qwen CLI}} \bigstrut\\\clineB{3-12}{2}
\textbf{Phase} & \multicolumn{1}{l}{\textbf{Failure category}}
  & \textit{w} & \textit{f} & \textit{w} & \textit{f}
  & \textit{w} & \textit{f} & \textit{w} & \textit{f}
  & \textit{w} & \textit{f} \bigstrut\\\clineB{3-12}{2}
\multicolumn{1}{c}{} & \multicolumn{1}{c}{}
  & \multicolumn{1}{c}{} & \multicolumn{1}{c}{} & \multicolumn{1}{c}{} & \multicolumn{1}{c}{}
  & \multicolumn{1}{c}{} & \multicolumn{1}{c}{} & \multicolumn{1}{c}{} & \multicolumn{1}{c}{}
  & \multicolumn{1}{c}{} & \multicolumn{1}{c}{} \\[-0.85em]\hlineB{2}

\multirow{4}{*}{\rotatebox{90}{\textbf{Build}}}
  & Dependency resolution     & \hm{35} & \hm{52} & \hm{16} & \hm{27} & \hm{11} & \hm{27} & \hm{40} & \hm{79} & \hm{18} & \hm{40} 
  \bigstrut[t]\\
  & Code compilation          & \hm{9} & \hm{17} & \hm{36} & \hm{48} & \hm{52} & \hm{44} & \hm{12} & \hm{1} & \hm{39} & \hm{37} \\
  & Project structure         & \hm{26} & \hm{8} & \hm{28} & \hm{7} & \hm{21} & \hm{9} & \hm{20} & \hm{4} & \hm{26} & \hm{6} \\
  & Maven plugin              & \hm{30} & \hm{23} & \hm{20} & \hm{17} & \hm{16} & \hm{20} & \hm{28} & \hm{16} & \hm{16} & \hm{16}\\
\hlineB{2}
\multicolumn{1}{c}{} & \multicolumn{1}{c}{}
  & \multicolumn{1}{c}{} & \multicolumn{1}{c}{} & \multicolumn{1}{c}{} & \multicolumn{1}{c}{}
  & \multicolumn{1}{c}{} & \multicolumn{1}{c}{} & \multicolumn{1}{c}{} & \multicolumn{1}{c}{}
  & \multicolumn{1}{c}{} & \multicolumn{1}{c}{} \\[-0.85em]\hlineB{2}

\multirow{5}{*}{\rotatebox{90}{\textbf{Deploy}}}
  & Resource / artifact       & \hm{28} & \hm{22} & \hm{32} & \hm{40} & \hm{40} & \hm{33} & \hm{21} & \hm{14} & \hm{40} & \hm{20} \bigstrut[t]\\
  & Config / startup          & \hm{6} & \hm{29} & \hm{16} & \hm{22} & \hm{13} & \hm{34} & \hm{21} & \hm{50} & \hm{13} & \hm{43} \\
  & Network / timeout         & \hm{17} & \hm{40} & \hm{8} & \hm{4} & \hm{13} & \hm{16} & \hm{11} & \hm{15} & \hm{20} & \hm{11} \\
  & Container exit            & \hm{11} & \hm{4} & \hm{8} & \hm{21} & \hm{0} & \hm{4} & \hm{7} & \hm{9} & \hm{7} & \hm{8} \\
  & DI / DB / class loading   & \hm{38} & \hm{4} & \hm{36} & \hm{12} & \hm{33} & \hm{13} & \hm{39} & \hm{12} & \hm{20} & \hm{17}\\
\hlineB{2}
\multicolumn{1}{c}{} & \multicolumn{1}{c}{}
  & \multicolumn{1}{c}{} & \multicolumn{1}{c}{} & \multicolumn{1}{c}{} & \multicolumn{1}{c}{}
  & \multicolumn{1}{c}{} & \multicolumn{1}{c}{} & \multicolumn{1}{c}{} & \multicolumn{1}{c}{}
  & \multicolumn{1}{c}{} & \multicolumn{1}{c}{} \\[-0.85em]\hlineB{2}
  
\multirow{4}{*}{\rotatebox{90}{\textbf{Test}}}
  & HTTP 404 / endpoint missing  & \hm{41} & \hm{26} & \hm{60} & \hm{15} & \hm{20} & \hm{28} & \hm{0} & \hm{31} & \hm{0} & \hm{37} \bigstrut[t]\\
  & HTTP 5xx / server error      & \hm{6} & \hm{0} & \hm{0} & \hm{0} & \hm{0} & \hm{0} & \hm{100} & \hm{4} & \hm{100} & \hm{10} \\
  & Assertion / content mismatch & \hm{35} & \hm{23} & \hm{20} & \hm{23} & \hm{40} & \hm{25} & \hm{0} & \hm{22} & \hm{0} & \hm{19} \\
  & Network / infrastructure     & \hm{18} & \hm{51} & \hm{20} & \hm{62} & \hm{40} & \hm{47} & \hm{0} & \hm{42} & \hm{0} & \hm{33} \\
\hlineB{2}
\end{tabular}%
}
\end{table*}

We analyze agent failures by execution phase (build, deploy, test) and by finer-grained failure modes within each phase. Experts first constructed a taxonomy by manually inspecting failed migrations. Two independent LLM annotators, Claude Opus-4.7 and GPT-5.5 (xhigh), then labeled each failed conversion using this fixed taxonomy, after which experts audited disagreements and adjudicated final labels. Inter-annotator agreement between the LLM annotators reached Cohen's $\kappa = 0.72$.

Table~\ref{tab:taxonomy-heatmap} summarizes the distribution of failure modes within each phase. Overall, failures extend beyond source-level translation errors and frequently arise from deployment, configuration, and behavioral inconsistencies that emerge only during runtime evaluation.

\bi
\item \textbf{Build-stage failures.} Common failures include dependency resolution, code compilation, and project-structure mismatches. Agents frequently leave stale module references, unresolved imports, or missing framework-specific build configuration.

\item \textbf{Deploy failures.} Deployment failures are dominated by build/launch mismatches, incorrect runtime configuration, and broken dependency wiring. These often reflect coordination failures between generated artifacts, runtime settings, and infrastructure rather than isolated code translation errors.

\item \textbf{Test failures.} Test failures commonly involve missing or incorrectly mapped endpoints, application reachability issues, and incorrect response content. Even when applications compile and deploy successfully, preserving behavioral equivalence across frameworks remains difficult.
\ei

\section{Limitations}
\label{sec:limitations}

\textbf{Evaluation Protocol.}~\scarfbench{} evaluates migrations through a sequential compile--deploy--test pipeline. As a result, candidates that correctly perform much of the source-level migration but fail due to dependency resolution, build configuration, or container startup are recorded at the earliest failing stage rather than evaluated for downstream correctness. We also evaluate each $(\text{agent}, \text{task})$ pair once at temperature~0. Larger sampling budgets or iterative repair could improve absolute success rates, so our results should be viewed as a compute-bounded estimate of current capability rather than an upper bound.

\textbf{Oracle Scope.}~
The behavioral oracle checks equivalence only at the observable boundary—HTTP routes, response payloads, UI flows, and persistent state. It does not capture internal issues such as race conditions, resource leaks, or non-functional regressions (e.g., latency or memory). We also do not directly score framework idiomaticity; such issues surface only indirectly through the failure taxonomy introduced in Section \ref{sec:failure-taxonomy}. While the oracles are expert-authored, we do not report formal coverage guarantees, so passing all tests does not ensure absence of defects.

\textbf{Annotation.}~
The failure taxonomy is induced and arbitrated by expert developers, but the LLM annotators used at scale are not separately calibrated against human labels (the reported $\kappa = 0.72$ is between LLM annotators only). The taxonomy is also not exhaustive and may miss failure modes arising in other agents, frameworks, or migration settings.

\textbf{Scope and Contamination.}~
\scarfbench{} focuses on enterprise Java migrations, and findings may not generalize to ecosystems such as .NET or Node.js. The benchmark applications are drawn from public repositories, including Eclipse Jakarta EE examples and \texttt{spring-petclinic}, which likely appear in frontier-model training data. We do not perform model-specific decontamination, so reported results may partially reflect memorization in addition to migration capability. Long-term reproducibility will also require ongoing maintenance as framework and tooling versions evolve.
\section{Conclusion}
\label{sec:conclusion}

We present \scarfbench{}, a benchmark for behavior-preserving cross-framework refactoring of enterprise Java applications. \scarfbench{} pairs 34 application families across Spring, Jakarta EE, and Quarkus into 102 expert-written framework variants and 204 directed migration tasks, each verified by a containerized compile-deploy-test harness against application-specific behavioral oracles. Across five state-of-the-art coding agents, the strongest run reached only $15.3\%$ aggregate test pass on focused-layer migrations and $12.2\%$ on whole applications; just one of the 204 tasks produced a fully behaviorally equivalent target. 

Our results show that current agents reliably translate framework APIs at the source level, yet rarely re-express application behavior through the target framework's runtime model. Failures are asymmetric, concentrating in Jakarta-targeted migrations and at coordination points spanning configuration, dependency wiring and deployment suggesting that progress on repository-level benchmarks does not transfer cleanly to system-level transformations that must execute in a target runtime.

\bibliographystyle{plainnat}
\bibliography{references}

\appendix

\onecolumn
\raggedbottom
\sloppy
\setlength{\emergencystretch}{3em}

\section{Benchmark dataset}
\label{app:dataset}

\scarfbench{} comprises 34 application families, each implemented in three
frameworks (Spring, Jakarta~EE, Quarkus), yielding 102 framework variants.
Each application is paired with a behavioral test suite expressed as a Gherkin
feature file; the harness compiles a smoke-test executable from the feature
file and runs it against the deployed migrated candidate. Tables~\ref{tab:dataset-focused}
and~\ref{tab:dataset-whole} list every application, its purpose, the number of
behavioral smoke tests, and the project size (median across the three
framework variants, reported as total source lines of code via
\texttt{tokei}\footnote{\url{https://github.com/XAMPPRocky/tokei}};
total includes Java, XML, Dockerfiles, web resources, and build shell
scripts).

\begin{table}[!bp]
\centering
\footnotesize
\setlength{\tabcolsep}{3pt}
\resizebox{\linewidth}{!}{%
\begin{threeparttable}
\begin{tabular}{@{}ll>{\raggedright\arraybackslash}p{5.5cm}rr@{}}
\toprule
\textbf{Layer} & \textbf{Application} & \textbf{Description} & \textbf{Tests} & \textbf{KLOC} \\
\midrule
\multirow{5}{*}{Business Domain}
 & \texttt{cart}            & Shopping cart with add/view/remove operations             & 8--14\tnote{a} & 0.82 \\
 & \texttt{converter}       & Currency conversion (USD $\to$ JPY, EUR)                  & 21    & 0.55 \\
 & \texttt{counter}         & Page-hit counter persisted across requests                & 9     & 0.55 \\
 & \texttt{helloservice}    & SOAP web service returning a greeting                     & 17    & 0.56 \\
 & \texttt{standalone}      & Minimal standalone REST greeting endpoint                 & 9     & 0.54 \\
\midrule
\multirow{7}{*}{Dependency Inj.}
 & \texttt{billpayment}     & Bill payments via CDI events and interceptors             & 14    & 0.84 \\
 & \texttt{decorators}      & String encoder wrapped by a CDI decorator                 & 12    & 0.76 \\
 & \texttt{encoder}         & Caesar cipher selectable via CDI alternatives             & 11    & 0.78 \\
 & \texttt{guessnumber}     & Number-guessing game with request-scoped CDI              & 9     & 0.80 \\
 & \texttt{producerfields}  & JPA to-do list via a CDI producer field                   & 8     & 0.80 \\
 & \texttt{producermethods} & Encoder selected at runtime via CDI producer method       & 12    & 0.77 \\
 & \texttt{simplegreeting}  & Greeting style chosen by a CDI qualifier                  & 7     & 0.65 \\
\midrule
\multirow{5}{*}{Infrastructure}
 & \texttt{concurrency-jobs}        & Job submission API with priority executors         & 8     & 0.70 \\
 & \texttt{concurrency-taskcreator} & Managed executors for immediate/delayed/periodic   & 6     & 0.79 \\
 & \texttt{ejb-async}               & Asynchronous email send via EJB \texttt{@Asynchronous} & 1--11\tnote{a} & 0.86 \\
 & \texttt{ejb-interceptor}         & EJB interceptor lowercasing greeting names         & 8     & 0.56 \\
 & \texttt{ejb-timersession}        & Programmatic and automatic EJB timers              & 6     & 0.55 \\
\midrule
\multirow{3}{*}{Persistence}
 & \texttt{address-book}    & Contact CRUD with field validation                        & 10    & 1.09 \\
 & \texttt{order}           & Orders with line items, parts, and vendors                & 7--11\tnote{a}  & 1.65 \\
 & \texttt{roster}          & Leagues/teams/players via JPA Criteria API                & 24--25\tnote{a} & 1.95 \\
\midrule
\multirow{9}{*}{Presentation}
 & \texttt{dukeetf}         & Async servlet streaming ETF ticks via long polling        & 4     & 0.63 \\
 & \texttt{dukeetf2}        & WebSocket variant of the ETF tick stream                  & 11    & 0.71 \\
 & \texttt{fileupload}      & Multipart file-upload servlet                             & 6     & 0.60 \\
 & \texttt{hello-servlet}   & Servlet returning a personalized greeting                 & 15    & 0.47 \\
 & \texttt{jaxrs-customer}  & Customer CRUD over JAX-RS with JPA                        & 10    & 1.02 \\
 & \texttt{jaxrs-hello}     & Minimal JAX-RS hello-world endpoint                       & 8     & 0.48 \\
 & \texttt{jaxrs-rsvp}      & RSVP event tracker over JAX-RS                            & 10    & 1.27 \\
 & \texttt{mood}            & Servlet rendering Duke's mood via a TimeOfDayFilter       & 7     & 0.64 \\
 & \texttt{websocketbot}    & WebSocket chat bot with rooms and broadcast               & 21    & 1.02 \\
\bottomrule
\end{tabular}
\begin{tablenotes}\footnotesize
\item[a] Test count varies across the three framework variants
  (Jakarta~/~Quarkus~/~Spring): \texttt{cart} 14/8/14, \texttt{ejb-async}
  11/1/11, \texttt{order} 11/7/11, \texttt{roster} 25/25/24. The
  \texttt{ejb-async} Quarkus count is degraded because Quarkus does not
  support EJB \texttt{@Asynchronous}; most scenarios are skipped.
\end{tablenotes}
\caption{\textbf{Focused tier --- 29 applications across five JSR-anchored
layers (87 variants).} Each application isolates a single architectural
concern and is shipped with a Gherkin feature file. The \emph{Tests} column
gives the number of behavioral smoke tests per framework variant; the
\emph{KLOC} column reports the median total source size across the three
variants.}
\label{tab:dataset-focused}
\end{threeparttable}
}
\end{table}

\begin{table}[!tbp]
\centering
\caption{\textbf{Whole-application tier --- 5 multi-layer applications
(15 variants).} These are realistic open-source enterprise Java applications
that combine persistence, presentation, infrastructure, and business-domain
logic. Columns as in Table~\ref{tab:dataset-focused}.}
\label{tab:dataset-whole}
\footnotesize
\setlength{\tabcolsep}{4pt}
\resizebox{\linewidth}{!}{%
\begin{threeparttable}
\begin{tabular}{@{}l>{\raggedright\arraybackslash}p{9.0cm}rr@{}}
\toprule
\textbf{Application} & \textbf{Description} & \textbf{Tests} & \textbf{KLOC} \\
\midrule
\texttt{cargotracker} & Domain-driven cargo-shipping lifecycle (Jakarta~EE reference application) & 34            & 25.72 \\
\texttt{coffee-shop}  & Microservices coffee-shop with asynchronous order routing                  & 9--11\tnote{a}  & 61.54 \\
\texttt{daytrader}    & Online stock-trading brokerage (IBM benchmark)                             & 20--30\tnote{a} & 14.31 \\
\texttt{petclinic}    & Veterinary clinic management (Spring reference application)               & 13--36\tnote{a,b} & 17.11\tnote{b} \\
\texttt{realworld}    & RealWorld blog/article/comment/follow REST API                             & 62            & 6.40 \\
\bottomrule
\end{tabular}
\begin{tablenotes}\footnotesize
\item[a] Test count varies across the three framework variants
  (Jakarta~/~Quarkus~/~Spring): \texttt{coffee-shop} 9/9/11,
  \texttt{daytrader} 21/30/20, \texttt{petclinic} 36/13/13.
\item[b] The \texttt{petclinic} variants exhibit substantial schema drift:
  the Jakarta variant is sourced from a richer upstream
  (\texttt{org.woehlke.jakartaee.petclinic}) carrying additional fields
  (\texttt{uuid}, \texttt{email}, \texttt{zipCode}, \ldots) that the Spring
  and Quarkus variants do not. KLOC ranges 11.88--22.11 across variants;
  median shown. We treat \texttt{petclinic} equivalence at the level of
  the shared API contract specified by the paired test suite. See
  \S\ref{sec:limitations}.
\end{tablenotes}
\end{threeparttable}
}
\end{table}

\subsection{Migration scale}
\label{app:migration-scale}

The two prior tables list every application; Table~\ref{tab:migration-scale}
quantifies the \emph{scope} of work each task asks of an agent, both
corpus-wide and per task.

\begin{table}[!tbp]
\centering
\caption{\textbf{\scarfbench{} migration scale.} Top: corpus-wide totals
covering files, source code, and behavioral specifications. Bottom: per-task
diff-and-context distributions. The per-task diff distribution is the
relevant comparator for SWE-bench-style benchmarks, where typical instances
are $<$100-line single-file patches; \scarfbench{} migrations span multiple
files and architectural layers by construction.}
\label{tab:migration-scale}
\footnotesize
\setlength{\tabcolsep}{6pt}
\resizebox{\linewidth}{!}{%
\begin{tabular}{lr}
\hlineB{2}
\multicolumn{2}{c}{\textsc{\textbf{Dataset totals}}} \bigstrut\\\hlineB{2}
Paired implementations (34 families $\times$ 3 frameworks) & 102 \bigstrut\\
Directed migration tasks (34 families $\times$ 6 pairs) & 204 \bigstrut\\
Java files & 1{,}946 \bigstrut\\
Java LOC & 151{,}471 \bigstrut\\
Build/config LOC (\texttt{pom.xml}, \texttt{web.xml}, \texttt{persistence.xml}, \texttt{beans.xml}) & $+$24{,}275 \bigstrut\\
\texttt{@Test} methods specifying behavior & 602 \bigstrut\\
Gherkin smoke-test scenarios (cf.\ Tables~\ref{tab:dataset-focused}--\ref{tab:dataset-whole}) & 1{,}389 \bigstrut\\\hlineB{2}
\multicolumn{2}{c}{\textsc{\textbf{Per-task scale}}} \bigstrut\\\hlineB{2}
Diff size, lines added$+$removed: median / mean / max\tnote{a} & 370 / 937 / 14{,}835 \bigstrut\\
``Only-in'' files per pair (created or deleted): mean / max & 11.4 / 66 \bigstrut\\
Source-side context the agent reads (LOC): median / mean / max & 360 / 1{,}485 / 19{,}836 \bigstrut\\
Total dataset edit volume across 204 tasks & ${\approx}$191 KLOC \bigstrut\\\hlineB{2}
\end{tabular}
}
\\[2pt]
{\scriptsize \textit{${}^{\textrm{a}}$ Max diff is DayTrader Spring$\leftrightarrow$Jakarta.}}
\end{table}

\setlength{\emergencystretch}{3em}

\section{Aggregate Leaderboard Visualization}
\label{app:aggregate-visualization}

\begin{figure*}[h!]
    \centering
    \includegraphics[width=\textwidth]{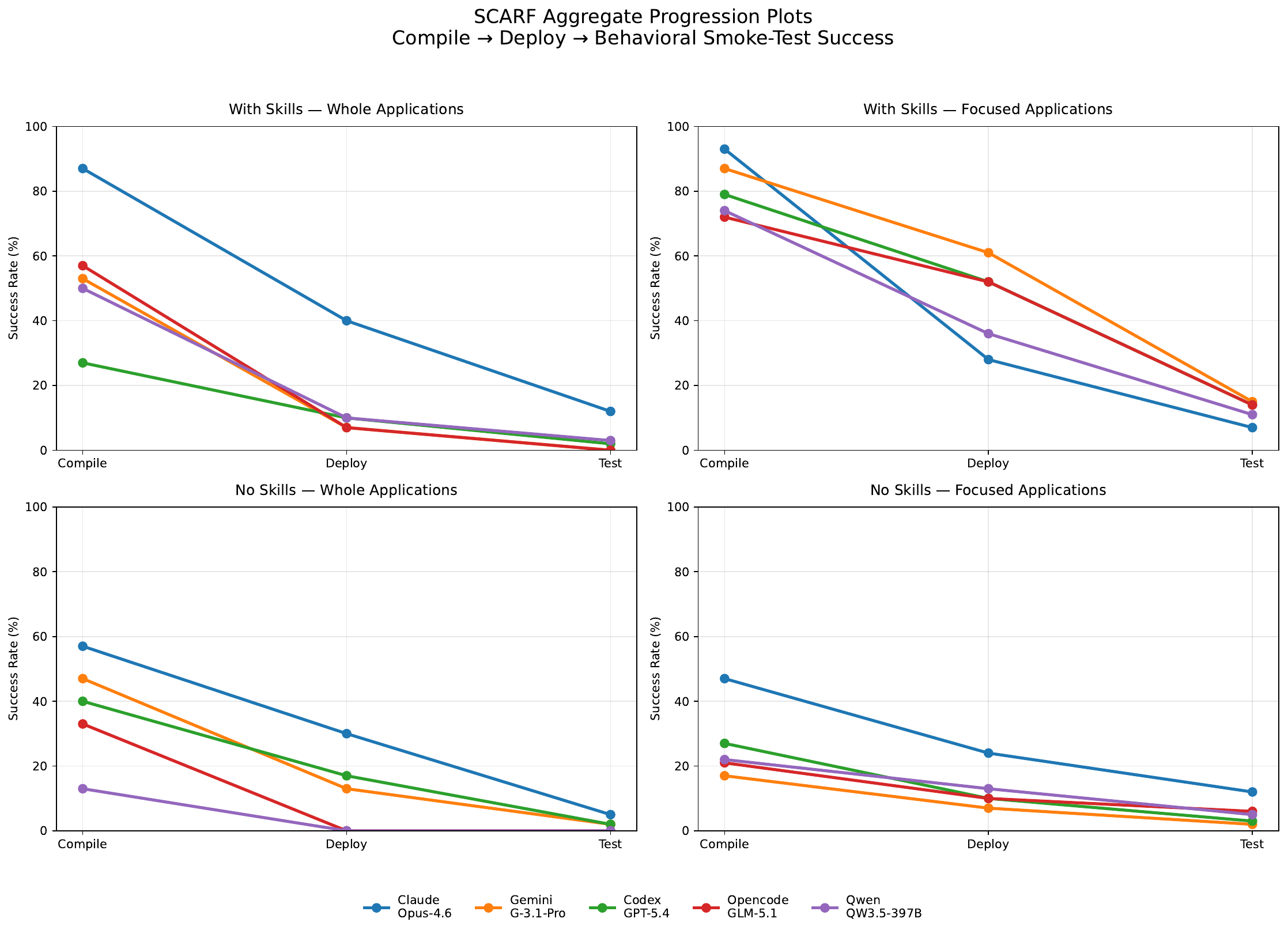}
    \caption{\textbf{Aggregate ScarfBench progression plots.} Each line traces
    the transition from compile to deploy to behavioral smoke-test success for
    a given harness--model pairing. Focused applications consistently achieve
    stronger end-to-end performance than whole applications. Skills-based
    prompting substantially improves focused-task deploy and behavioral success,
    particularly for Gemini CLI, while whole-application behavioral correctness
    remains limited across most harnesses.}
    \label{fig:aggregate-progression}
\end{figure*}

Figure~\ref{fig:aggregate-progression} visualizes the aggregate SCARF
leaderboard from Table~\ref{tab:scarf-leaderboard-aggregate} as a three-stage
migration pipeline consisting of compile, deploy, and behavioral smoke-test
success rates. The figure compares skills-enabled and no-skills prompting
across focused and whole-application settings.

The progression plots highlight several trends. First, compile success
consistently exceeds deploy and behavioral smoke-test success, indicating that
successful builds do not necessarily translate to correct runtime behavior.
This gap is especially pronounced for whole applications, where several
harnesses achieve moderate-to-high compile success but substantially lower
behavioral correctness. Second, focused applications achieve stronger
end-to-end performance than whole applications, particularly under
skills-enabled prompting. Third, skills-based prompting produces the largest
gains for focused-task migrations, most notably for Gemini CLI, which improves
focused deploy success from $7\%$ to $61\%$. Finally, Claude Code exhibits the
strongest whole-application performance overall, reaching $87\%$ compile,
$40\%$ deploy, and $12\%$ behavioral smoke-test success, while Gemini CLI
achieves the strongest focused-task deploy and smoke-test success rates.

%

\section{Leaderboard Breakdown}
\label{app:leaderboard-breakdown}

\paragraph{Per-direction leaderboard analysis.}
Tables~\ref{tab:scarf-leaderboard-skills}
and~\ref{tab:scarf-leaderboard-no-skills} provide a directional breakdown of
migration performance across the six framework transformations. Several trends
emerge across both focused and whole-application settings.

\begin{table}[h!]
\caption{\textbf{SCARF leaderboard --- per-direction breakdown, with skills.} Pass rates (\%) at the compile (\textit{c}), run/deploy (\textit{r}), and behavioral-test (\textit{t}) gates, across five agent harnesses and six source$\,\to\,$target framework migrations, evaluated with per-task skill prompts. Within each sub-table, the \textit{whole} half evaluates the multi-layer applications ($n{=}5$ for \textit{c}/\textit{r}); the \textit{focused} half evaluates the per-layer single-concern apps ($n{=}29$ for \textit{c}/\textit{r}). The \textit{t} column totals passed smoke tests over expected smoke tests separately within each direction and tier; rows with blank test output, unknown deployment, failed deployment, or a missing deploy-outcome record contribute zero passed tests.}
\label{tab:scarf-leaderboard-skills}
\scriptsize
\setlength{\tabcolsep}{3pt}

\resizebox{0.449\linewidth}{!}{%
\begin{tabular}{llrrrV{2}rrr}
\clineB{3-8}{2}
\multicolumn{2}{c}{} & \multicolumn{6}{c}{\textsc{\textbf{$\mathcal{Q}$ $\to$ $\mathcal{J}$}}} \bigstrut\\\clineB{3-8}{2}
\multicolumn{2}{c}{} &
  \multicolumn{3}{cV{2}}{\textit{whole}} &
  \multicolumn{3}{c}{\textit{focused}} \bigstrut\\\clineB{3-8}{2}
Harness & Model &
  {c~(\%)} & {r~(\%)} & {t~(\%)} &
  {c~(\%)} & {r~(\%)} & {t~(\%)} \bigstrut\\\clineB{3-8}{2}
       &              & \multicolumn{1}{c}{} & \multicolumn{1}{c}{} & \multicolumn{1}{c}{} & \multicolumn{1}{c}{} & \multicolumn{1}{c}{} & \multicolumn{1}{c}{} \bigstrut\\[-1.4em]\hlineB{2}
Claude Code & Opus-4.6   & {\color{white}0}20                                               & {\color{white}0}0                                                & {\color{white}0}0                                                & {\color{white}0}59                                               & {\color{white}0}0                                                & {\color{white}0}0                                                \bigstrut\\
Gemini CLI  & G-3.1-Pro  & {\color{white}0}20                                               & {\color{white}0}0                                                & {\color{white}0}0                                                & \cellcolor{blue!18}{\bfseries{\color{blue!18}0}79}               & \cellcolor{blue!18}{\bfseries{\color{blue!18}0}66}               & \cellcolor{blue!18}{\bfseries{\color{blue!18}0}8}                \bigstrut\\
Codex       & GPT-5.4    & {\color{white}0}20                                               & \cellcolor{blue!18}{\bfseries{\color{blue!18}0}20}               & \cellcolor{blue!18}{\bfseries{\color{blue!18}0}2}                & {\color{white}0}59                                               & {\color{white}0}48                                               & {\color{white}0}3                                                \bigstrut\\
Opencode    & GLM-5.1    & {\color{white}0}0                                                & {\color{white}0}0                                                & {\color{white}0}0                                                & {\color{white}0}21                                               & {\color{white}0}17                                               & {\color{white}0}0                                                \bigstrut\\
Qwen CLI    & QW3.5-397B  & \cellcolor{blue!18}{\bfseries{\color{blue!18}0}60}               & \cellcolor{blue!18}{\bfseries{\color{blue!18}0}20}               & {\color{white}0}0                                                & {\color{white}0}24                                               & {\color{white}0}17                                               & {\color{white}0}3                                                \bigstrut\\\hlineB{2}
\end{tabular}%
}
\resizebox{0.27\linewidth}{!}{%
\begin{tabular}{rrrV{2}rrr}
\clineB{1-6}{2}
\multicolumn{6}{c}{\textsc{\textbf{$\mathcal{J}$ $\to$ $\mathcal{S}$}}} \bigstrut\\\clineB{1-6}{2}
  \multicolumn{3}{cV{2}}{\textit{whole}} &
  \multicolumn{3}{c}{\textit{focused}} \bigstrut\\\clineB{1-6}{2}
  {c~(\%)} & {r~(\%)} & {t~(\%)} &
  {c~(\%)} & {r~(\%)} & {t~(\%)} \bigstrut\\\clineB{1-6}{2}
\multicolumn{1}{c}{} & \multicolumn{1}{c}{} & \multicolumn{1}{c}{} & \multicolumn{1}{c}{} & \multicolumn{1}{c}{} & \multicolumn{1}{c}{} \bigstrut\\[-1.4em]\hlineB{2}
\cellcolor{blue!18}\textbf{100}                                  & \cellcolor{blue!18}{\bfseries{\color{blue!18}0}60}               & \cellcolor{blue!18}{\bfseries{\color{blue!18}0}4}                & {\color{white}0}97                                               & \cellcolor{blue!18}{\bfseries{\color{blue!18}0}52}               & {\color{white}0}2                                                \bigstrut\\
{\color{white}0}80                                               & {\color{white}0}0                                                & {\color{white}0}0                                                & \cellcolor{blue!18}\textbf{100}                                  & {\color{white}0}31                                               & {\color{white}0}0                                                \bigstrut\\
{\color{white}0}40                                               & {\color{white}0}0                                                & {\color{white}0}0                                                & {\color{white}0}90                                               & {\color{white}0}45                                               & {\color{white}0}2                                                \bigstrut\\
\cellcolor{blue!18}\textbf{100}                                  & {\color{white}0}0                                                & {\color{white}0}0                                                & {\color{white}0}97                                               & \cellcolor{blue!18}{\bfseries{\color{blue!18}0}52}               & \cellcolor{blue!18}{\bfseries{\color{blue!18}0}6}                \bigstrut\\
{\color{white}0}60                                               & {\color{white}0}0                                                & {\color{white}0}0                                                & {\color{white}0}90                                               & {\color{white}0}34                                               & {\color{white}0}5                                                \bigstrut\\\hlineB{2}
\end{tabular}%
}
\resizebox{0.27\linewidth}{!}{%
\begin{tabular}{rrrV{2}rrr}
\clineB{1-6}{2}
\multicolumn{6}{c}{\textsc{\textbf{$\mathcal{J}$ $\to$ $\mathcal{Q}$}}} \bigstrut\\\clineB{1-6}{2}
  \multicolumn{3}{cV{2}}{\textit{whole}} &
  \multicolumn{3}{c}{\textit{focused}} \bigstrut\\\clineB{1-6}{2}
  {c~(\%)} & {r~(\%)} & {t~(\%)} &
  {c~(\%)} & {r~(\%)} & {t~(\%)} \bigstrut\\\clineB{1-6}{2}
\multicolumn{1}{c}{} & \multicolumn{1}{c}{} & \multicolumn{1}{c}{} & \multicolumn{1}{c}{} & \multicolumn{1}{c}{} & \multicolumn{1}{c}{} \bigstrut\\[-1.4em]\hlineB{2}
{\color{white}0}40                                               & \cellcolor{blue!18}{\bfseries{\color{blue!18}0}20}               & {\color{white}0}3                                                & \cellcolor{blue!18}{\bfseries{\color{blue!18}0}93}               & {\color{white}0}52                                               & {\color{white}0}2                                                \bigstrut\\
\cellcolor{blue!18}{\bfseries{\color{blue!18}0}80}               & {\color{white}0}0                                                & {\color{white}0}0                                                & {\color{white}0}86                                               & \cellcolor{blue!18}{\bfseries{\color{blue!18}0}55}               & \cellcolor{blue!18}{\bfseries{\color{blue!18}0}6}                \bigstrut\\
{\color{white}0}40                                               & \cellcolor{blue!18}{\bfseries{\color{blue!18}0}20}               & {\color{white}0}0                                                & {\color{white}0}79                                               & {\color{white}0}28                                               & {\color{white}0}0                                                \bigstrut\\
{\color{white}0}60                                               & {\color{white}0}0                                                & {\color{white}0}0                                                & \cellcolor{blue!18}{\bfseries{\color{blue!18}0}93}               & {\color{white}0}52                                               & \cellcolor{blue!18}{\bfseries{\color{blue!18}0}6}                \bigstrut\\
{\color{white}0}40                                               & \cellcolor{blue!18}{\bfseries{\color{blue!18}0}20}               & \cellcolor{blue!18}{\bfseries{\color{blue!18}0}8}                & {\color{white}0}76                                               & {\color{white}0}31                                               & {\color{white}0}2                                                \bigstrut\\\hlineB{2}
\end{tabular}%
}

\vspace{6pt}

\resizebox{0.448\linewidth}{!}{%
\begin{tabular}{llrrrV{2}rrr}
\clineB{3-8}{2}
\multicolumn{2}{c}{} & \multicolumn{6}{c}{\textsc{\textbf{$\mathcal{S}$ $\to$ $\mathcal{J}$}}} \bigstrut\\\clineB{3-8}{2}
\multicolumn{2}{c}{} &
  \multicolumn{3}{cV{2}}{\textit{whole}} &
  \multicolumn{3}{c}{\textit{focused}} \bigstrut\\\clineB{3-8}{2}
Harness & Model &
  {c~(\%)} & {r~(\%)} & {t~(\%)} &
  {c~(\%)} & {r~(\%)} & {t~(\%)} \bigstrut\\\clineB{3-8}{2}
       &              & \multicolumn{1}{c}{} & \multicolumn{1}{c}{} & \multicolumn{1}{c}{} & \multicolumn{1}{c}{} & \multicolumn{1}{c}{} & \multicolumn{1}{c}{} \bigstrut\\[-1.4em]\hlineB{2}
Claude Code & Opus-4.6   & \cellcolor{blue!18}{\bfseries{\color{blue!18}0}60}               & \cellcolor{blue!18}{\bfseries{\color{blue!18}0}60}               & \cellcolor{blue!18}{\bfseries{\color{blue!18}0}8}                & {\color{white}0}59                                               & {\color{white}0}0                                                & {\color{white}0}0                                                \bigstrut\\
Gemini CLI  & G-3.1-Pro  & {\color{white}0}20                                               & {\color{white}0}20                                               & {\color{white}0}1                                                & \cellcolor{blue!18}{\bfseries{\color{blue!18}0}72}               & \cellcolor{blue!18}{\bfseries{\color{blue!18}0}66}               & {\color{white}0}2                                                \bigstrut\\
Codex       & GPT-5.4    & {\color{white}0}0                                                & {\color{white}0}0                                                & {\color{white}0}0                                                & {\color{white}0}55                                               & {\color{white}0}55                                               & \cellcolor{blue!18}{\bfseries{\color{blue!18}0}6}                \bigstrut\\
Opencode    & GLM-5.1    & {\color{white}0}0                                                & {\color{white}0}0                                                & {\color{white}0}0                                                & {\color{white}0}24                                               & {\color{white}0}21                                               & {\color{white}0}2                                                \bigstrut\\
Qwen CLI    & QW3.5-397B  & {\color{white}0}20                                               & {\color{white}0}0                                                & {\color{white}0}0                                                & {\color{white}0}52                                               & {\color{white}0}38                                               & {\color{white}0}2                                                \bigstrut\\\hlineB{2}
\end{tabular}%
}
\resizebox{0.27\linewidth}{!}{%
\begin{tabular}{rrrV{2}rrr}
\clineB{1-6}{2}
\multicolumn{6}{c}{\textsc{\textbf{$\mathcal{Q}$ $\to$ $\mathcal{S}$}}} \bigstrut\\\clineB{1-6}{2}
  \multicolumn{3}{cV{2}}{\textit{whole}} &
  \multicolumn{3}{c}{\textit{focused}} \bigstrut\\\clineB{1-6}{2}
  {c~(\%)} & {r~(\%)} & {t~(\%)} &
  {c~(\%)} & {r~(\%)} & {t~(\%)} \bigstrut\\\clineB{1-6}{2}
\multicolumn{1}{c}{} & \multicolumn{1}{c}{} & \multicolumn{1}{c}{} & \multicolumn{1}{c}{} & \multicolumn{1}{c}{} & \multicolumn{1}{c}{} \bigstrut\\[-1.4em]\hlineB{2}
{\color{white}0}60                                               & \cellcolor{blue!18}{\bfseries{\color{blue!18}0}20}               & \cellcolor{blue!18}{\bfseries{\color{blue!18}0}9}                & {\color{white}0}52                                               & {\color{white}0}41                                               & {\color{white}0}19                                               \bigstrut\\
\cellcolor{blue!18}\textbf{100}                                  & \cellcolor{blue!18}{\bfseries{\color{blue!18}0}20}               & {\color{white}0}1                                                & \cellcolor{blue!18}\textbf{100}                                  & {\color{white}0}72                                               & {\color{white}0}28                                               \bigstrut\\
{\color{white}0}60                                               & \cellcolor{blue!18}{\bfseries{\color{blue!18}0}20}               & {\color{white}0}8                                                & \cellcolor{blue!18}\textbf{100}                                  & {\color{white}0}69                                               & {\color{white}0}39                                               \bigstrut\\
\cellcolor{blue!18}\textbf{100}                                  & \cellcolor{blue!18}{\bfseries{\color{blue!18}0}20}               & {\color{white}0}0                                                & \cellcolor{blue!18}\textbf{100}                                  & \cellcolor{blue!18}{\bfseries{\color{blue!18}0}86}               & \cellcolor{blue!18}{\bfseries{\color{blue!18}0}43}               \bigstrut\\
{\color{white}0}60                                               & \cellcolor{blue!18}{\bfseries{\color{blue!18}0}20}               & \cellcolor{blue!18}{\bfseries{\color{blue!18}0}9}                & {\color{white}0}90                                               & {\color{white}0}48                                               & {\color{white}0}20                                               \bigstrut\\\hlineB{2}
\end{tabular}%
}
\resizebox{0.27\linewidth}{!}{%
\begin{tabular}{rrrV{2}rrr}
\clineB{1-6}{2}
\multicolumn{6}{c}{\textsc{\textbf{$\mathcal{S}$ $\to$ $\mathcal{Q}$}}} \bigstrut\\\clineB{1-6}{2}
  \multicolumn{3}{cV{2}}{\textit{whole}} &
  \multicolumn{3}{c}{\textit{focused}} \bigstrut\\\clineB{1-6}{2}
  {c~(\%)} & {r~(\%)} & {t~(\%)} &
  {c~(\%)} & {r~(\%)} & {t~(\%)} \bigstrut\\\clineB{1-6}{2}
\multicolumn{1}{c}{} & \multicolumn{1}{c}{} & \multicolumn{1}{c}{} & \multicolumn{1}{c}{} & \multicolumn{1}{c}{} & \multicolumn{1}{c}{} \bigstrut\\[-1.4em]\hlineB{2}
\cellcolor{blue!18}\textbf{100}                                  & \cellcolor{blue!18}{\bfseries{\color{blue!18}0}80}               & \cellcolor{blue!18}{\bfseries{\color{blue!18}0}50}               & {\color{white}0}52                                               & {\color{white}0}24                                               & {\color{white}0}21                                               \bigstrut\\
{\color{white}0}20                                               & {\color{white}0}0                                                & {\color{white}0}0                                                & {\color{white}0}86                                               & {\color{white}0}76                                               & \cellcolor{blue!18}{\bfseries{\color{blue!18}0}49}               \bigstrut\\
{\color{white}0}0                                                & {\color{white}0}0                                                & {\color{white}0}0                                                & {\color{white}0}93                                               & {\color{white}0}66                                               & {\color{white}0}33                                               \bigstrut\\
{\color{white}0}80                                               & {\color{white}0}20                                               & {\color{white}0}0                                                & \cellcolor{blue!18}\textbf{100}                                  & \cellcolor{blue!18}{\bfseries{\color{blue!18}0}86}               & {\color{white}0}29                                               \bigstrut\\
{\color{white}0}20                                               & {\color{white}0}0                                                & {\color{white}0}0                                                & {\color{white}0}79                                               & {\color{white}0}45                                               & {\color{white}0}33                                               \bigstrut\\\hlineB{2}
\end{tabular}%
}

\tiny{\textit{Note: \colorbox{blue!18}{\textbf{$bold$}} marks the per-column max within each direction$\times$tier; all-zero columns are left unbolded.}}
\end{table}


\begin{table}[h!]
\caption{\textbf{SCARF leaderboard --- per-direction breakdown, no skills.} Pass rates (\%) at the compile (\textit{c}), run/deploy (\textit{r}), and behavioral-test (\textit{t}) gates, across five agent harnesses and six source$\,\to\,$target framework migrations, evaluated without per-task skill prompts. Within each sub-table, the \textit{whole} half evaluates the multi-layer applications ($n{=}5$ for \textit{c}/\textit{r}); the \textit{focused} half evaluates the per-layer single-concern apps ($n{=}29$ for \textit{c}/\textit{r}). The \textit{t} column totals passed smoke tests over expected smoke tests separately within each direction and tier; rows with blank test output, unknown deployment, failed deployment, or a missing deploy-outcome record contribute zero passed tests.}
\label{tab:scarf-leaderboard-no-skills}
\scriptsize
\setlength{\tabcolsep}{3pt}

\resizebox{0.448\linewidth}{!}{%
\begin{tabular}{llrrrV{2}rrr}
\clineB{3-8}{2}
\multicolumn{2}{c}{} & \multicolumn{6}{c}{\textsc{\textbf{$\mathcal{Q}$ $\to$ $\mathcal{J}$}}} \bigstrut\\\clineB{3-8}{2}
\multicolumn{2}{c}{} &
  \multicolumn{3}{cV{2}}{\textit{whole}} &
  \multicolumn{3}{c}{\textit{focused}} \bigstrut\\\clineB{3-8}{2}
Harness & Model &
  {c~(\%)} & {r~(\%)} & {t~(\%)} &
  {c~(\%)} & {r~(\%)} & {t~(\%)} \bigstrut\\\clineB{3-8}{2}
       &              & \multicolumn{1}{c}{} & \multicolumn{1}{c}{} & \multicolumn{1}{c}{} & \multicolumn{1}{c}{} & \multicolumn{1}{c}{} & \multicolumn{1}{c}{} \bigstrut\\[-1.4em]\hlineB{2}
Claude Code & Opus-4.6   & {\color{white}0}20                                               & {\color{white}0}0                                                & {\color{white}0}0                                                & {\color{white}0}14                                               & {\color{white}0}0                                                & {\color{white}0}0                                                \bigstrut\\
Gemini CLI  & G-3.1-Pro  & \cellcolor{blue!18}{\bfseries{\color{blue!18}0}40}               & {\color{white}0}0                                                & {\color{white}0}0                                                & \cellcolor{blue!18}{\bfseries{\color{blue!18}0}24}               & \cellcolor{blue!18}{\bfseries{\color{blue!18}0}7}                & {\color{white}0}0                                                \bigstrut\\
Codex       & GPT-5.4    & {\color{white}0}20                                               & {\color{white}0}0                                                & {\color{white}0}0                                                & {\color{white}0}21                                               & \cellcolor{blue!18}{\bfseries{\color{blue!18}0}7}                & {\color{white}0}0                                                \bigstrut\\
Opencode    & GLM-5.1    & {\color{white}0}20                                               & {\color{white}0}0                                                & {\color{white}0}0                                                & {\color{white}0}14                                               & \cellcolor{blue!18}{\bfseries{\color{blue!18}0}7}                & \cellcolor{blue!18}{\bfseries{\color{blue!18}0}2}                \bigstrut\\
Qwen CLI    & QW3.5-397B  & {\color{white}0}20                                               & {\color{white}0}0                                                & {\color{white}0}0                                                & {\color{white}0}21                                               & \cellcolor{blue!18}{\bfseries{\color{blue!18}0}7}                & {\color{white}0}0                                                \bigstrut\\\hlineB{2}
\end{tabular}%
}
\resizebox{0.27\linewidth}{!}{%
\begin{tabular}{rrrV{2}rrr}
\clineB{1-6}{2}
\multicolumn{6}{c}{\textsc{\textbf{$\mathcal{J}$ $\to$ $\mathcal{S}$}}} \bigstrut\\\clineB{1-6}{2}
  \multicolumn{3}{cV{2}}{\textit{whole}} &
  \multicolumn{3}{c}{\textit{focused}} \bigstrut\\\clineB{1-6}{2}
  {c~(\%)} & {r~(\%)} & {t~(\%)} &
  {c~(\%)} & {r~(\%)} & {t~(\%)} \bigstrut\\\clineB{1-6}{2}
\multicolumn{1}{c}{} & \multicolumn{1}{c}{} & \multicolumn{1}{c}{} & \multicolumn{1}{c}{} & \multicolumn{1}{c}{} & \multicolumn{1}{c}{} \bigstrut\\[-1.4em]\hlineB{2}
\cellcolor{blue!18}{\bfseries{\color{blue!18}0}80}               & \cellcolor{blue!18}{\bfseries{\color{blue!18}0}40}               & \cellcolor{blue!18}{\bfseries{\color{blue!18}0}5}                & \cellcolor{blue!18}{\bfseries{\color{blue!18}0}83}               & \cellcolor{blue!18}{\bfseries{\color{blue!18}0}24}               & \cellcolor{blue!18}{\bfseries{\color{blue!18}0}12}               \bigstrut\\
{\color{white}0}60                                               & {\color{white}0}20                                               & {\color{white}0}2                                                & {\color{white}0}21                                               & {\color{white}0}14                                               & {\color{white}0}9                                                \bigstrut\\
{\color{white}0}40                                               & {\color{white}0}20                                               & {\color{white}0}2                                                & {\color{white}0}31                                               & {\color{white}0}7                                                & {\color{white}0}4                                                \bigstrut\\
{\color{white}0}40                                               & {\color{white}0}0                                                & {\color{white}0}0                                                & {\color{white}0}14                                               & {\color{white}0}3                                                & {\color{white}0}4                                                \bigstrut\\
{\color{white}0}20                                               & {\color{white}0}0                                                & {\color{white}0}0                                                & {\color{white}0}28                                               & {\color{white}0}21                                               & {\color{white}0}7                                                \bigstrut\\\hlineB{2}
\end{tabular}%
}
\resizebox{0.27\linewidth}{!}{%
\begin{tabular}{rrrV{2}rrr}
\clineB{1-6}{2}
\multicolumn{6}{c}{\textsc{\textbf{$\mathcal{J}$ $\to$ $\mathcal{Q}$}}} \bigstrut\\\clineB{1-6}{2}
  \multicolumn{3}{cV{2}}{\textit{whole}} &
  \multicolumn{3}{c}{\textit{focused}} \bigstrut\\\clineB{1-6}{2}
  {c~(\%)} & {r~(\%)} & {t~(\%)} &
  {c~(\%)} & {r~(\%)} & {t~(\%)} \bigstrut\\\clineB{1-6}{2}
\multicolumn{1}{c}{} & \multicolumn{1}{c}{} & \multicolumn{1}{c}{} & \multicolumn{1}{c}{} & \multicolumn{1}{c}{} & \multicolumn{1}{c}{} \bigstrut\\[-1.4em]\hlineB{2}
{\color{white}0}80                                               & \cellcolor{blue!18}{\bfseries{\color{blue!18}0}40}               & \cellcolor{blue!18}{\bfseries{\color{blue!18}0}5}                & \cellcolor{blue!18}{\bfseries{\color{blue!18}0}83}               & \cellcolor{blue!18}{\bfseries{\color{blue!18}0}48}               & \cellcolor{blue!18}{\bfseries{\color{blue!18}0}15}               \bigstrut\\
{\color{white}0}60                                               & {\color{white}0}0                                                & {\color{white}0}0                                                & {\color{white}0}28                                               & {\color{white}0}3                                                & {\color{white}0}0                                                \bigstrut\\
\cellcolor{blue!18}\textbf{100}                                  & \cellcolor{blue!18}{\bfseries{\color{blue!18}0}40}               & {\color{white}0}3                                                & {\color{white}0}31                                               & {\color{white}0}10                                               & {\color{white}0}9                                                \bigstrut\\
{\color{white}0}40                                               & {\color{white}0}0                                                & {\color{white}0}0                                                & {\color{white}0}34                                               & {\color{white}0}24                                               & {\color{white}0}12                                               \bigstrut\\
{\color{white}0}0                                                & {\color{white}0}0                                                & {\color{white}0}0                                                & {\color{white}0}24                                               & {\color{white}0}17                                               & {\color{white}0}7                                                \bigstrut\\\hlineB{2}
\end{tabular}%
}

\vspace{6pt}

\resizebox{0.448\linewidth}{!}{%
\begin{tabular}{llrrrV{2}rrr}
\clineB{3-8}{2}
\multicolumn{2}{c}{} & \multicolumn{6}{c}{\textsc{\textbf{$\mathcal{S}$ $\to$ $\mathcal{J}$}}} \bigstrut\\\clineB{3-8}{2}
\multicolumn{2}{c}{} &
  \multicolumn{3}{cV{2}}{\textit{whole}} &
  \multicolumn{3}{c}{\textit{focused}} \bigstrut\\\clineB{3-8}{2}
Harness & Model &
  {c~(\%)} & {r~(\%)} & {t~(\%)} &
  {c~(\%)} & {r~(\%)} & {t~(\%)} \bigstrut\\\clineB{3-8}{2}
       &              & \multicolumn{1}{c}{} & \multicolumn{1}{c}{} & \multicolumn{1}{c}{} & \multicolumn{1}{c}{} & \multicolumn{1}{c}{} & \multicolumn{1}{c}{} \bigstrut\\[-1.4em]\hlineB{2}
Claude Code & Opus-4.6   & {\color{white}0}0                                                & {\color{white}0}0                                                & {\color{white}0}0                                                & {\color{white}0}21                                               & {\color{white}0}10                                               & \cellcolor{blue!18}{\bfseries{\color{blue!18}0}2}                \bigstrut\\
Gemini CLI  & G-3.1-Pro  & \cellcolor{blue!18}{\bfseries{\color{blue!18}0}20}               & {\color{white}0}0                                                & {\color{white}0}0                                                & {\color{white}0}10                                               & {\color{white}0}7                                                & {\color{white}0}0                                                \bigstrut\\
Codex       & GPT-5.4    & \cellcolor{blue!18}{\bfseries{\color{blue!18}0}20}               & {\color{white}0}0                                                & {\color{white}0}0                                                & \cellcolor{blue!18}{\bfseries{\color{blue!18}0}31}               & \cellcolor{blue!18}{\bfseries{\color{blue!18}0}17}               & {\color{white}0}0                                                \bigstrut\\
Opencode    & GLM-5.1    & \cellcolor{blue!18}{\bfseries{\color{blue!18}0}20}               & {\color{white}0}0                                                & {\color{white}0}0                                                & {\color{white}0}17                                               & {\color{white}0}0                                                & {\color{white}0}0                                                \bigstrut\\
Qwen CLI    & QW3.5-397B  & {\color{white}0}0                                                & {\color{white}0}0                                                & {\color{white}0}0                                                & {\color{white}0}24                                               & {\color{white}0}7                                                & {\color{white}0}0                                                \bigstrut\\\hlineB{2}
\end{tabular}%
}
\resizebox{0.27\linewidth}{!}{%
\begin{tabular}{rrrV{2}rrr}
\clineB{1-6}{2}
\multicolumn{6}{c}{\textsc{\textbf{$\mathcal{Q}$ $\to$ $\mathcal{S}$}}} \bigstrut\\\clineB{1-6}{2}
  \multicolumn{3}{cV{2}}{\textit{whole}} &
  \multicolumn{3}{c}{\textit{focused}} \bigstrut\\\clineB{1-6}{2}
  {c~(\%)} & {r~(\%)} & {t~(\%)} &
  {c~(\%)} & {r~(\%)} & {t~(\%)} \bigstrut\\\clineB{1-6}{2}
\multicolumn{1}{c}{} & \multicolumn{1}{c}{} & \multicolumn{1}{c}{} & \multicolumn{1}{c}{} & \multicolumn{1}{c}{} & \multicolumn{1}{c}{} \bigstrut\\[-1.4em]\hlineB{2}
\cellcolor{blue!18}{\bfseries{\color{blue!18}0}80}               & \cellcolor{blue!18}{\bfseries{\color{blue!18}0}60}               & \cellcolor{blue!18}{\bfseries{\color{blue!18}0}11}               & \cellcolor{blue!18}{\bfseries{\color{blue!18}0}41}               & \cellcolor{blue!18}{\bfseries{\color{blue!18}0}24}               & \cellcolor{blue!18}{\bfseries{\color{blue!18}0}18}               \bigstrut\\
{\color{white}0}60                                               & {\color{white}0}40                                               & {\color{white}0}3                                                & {\color{white}0}3                                                & {\color{white}0}0                                                & {\color{white}0}0                                                \bigstrut\\
{\color{white}0}60                                               & {\color{white}0}40                                               & {\color{white}0}9                                                & \cellcolor{blue!18}{\bfseries{\color{blue!18}0}41}               & {\color{white}0}14                                               & {\color{white}0}4                                                \bigstrut\\
{\color{white}0}40                                               & {\color{white}0}0                                                & {\color{white}0}0                                                & {\color{white}0}14                                               & {\color{white}0}10                                               & {\color{white}0}8                                                \bigstrut\\
{\color{white}0}20                                               & {\color{white}0}0                                                & {\color{white}0}0                                                & {\color{white}0}7                                                & {\color{white}0}7                                                & {\color{white}0}3                                                \bigstrut\\\hlineB{2}
\end{tabular}%
}
\resizebox{0.27\linewidth}{!}{%
\begin{tabular}{rrrV{2}rrr}
\clineB{1-6}{2}
\multicolumn{6}{c}{\textsc{\textbf{$\mathcal{S}$ $\to$ $\mathcal{Q}$}}} \bigstrut\\\clineB{1-6}{2}
  \multicolumn{3}{cV{2}}{\textit{whole}} &
  \multicolumn{3}{c}{\textit{focused}} \bigstrut\\\clineB{1-6}{2}
  {c~(\%)} & {r~(\%)} & {t~(\%)} &
  {c~(\%)} & {r~(\%)} & {t~(\%)} \bigstrut\\\clineB{1-6}{2}
\multicolumn{1}{c}{} & \multicolumn{1}{c}{} & \multicolumn{1}{c}{} & \multicolumn{1}{c}{} & \multicolumn{1}{c}{} & \multicolumn{1}{c}{} \bigstrut\\[-1.4em]\hlineB{2}
\cellcolor{blue!18}{\bfseries{\color{blue!18}0}80}               & \cellcolor{blue!18}{\bfseries{\color{blue!18}0}40}               & \cellcolor{blue!18}{\bfseries{\color{blue!18}0}9}                & \cellcolor{blue!18}{\bfseries{\color{blue!18}0}41}               & \cellcolor{blue!18}{\bfseries{\color{blue!18}0}38}               & \cellcolor{blue!18}{\bfseries{\color{blue!18}0}23}               \bigstrut\\
{\color{white}0}40                                               & {\color{white}0}20                                               & {\color{white}0}8                                                & {\color{white}0}14                                               & {\color{white}0}10                                               & {\color{white}0}2                                                \bigstrut\\
{\color{white}0}0                                                & {\color{white}0}0                                                & {\color{white}0}0                                                & {\color{white}0}7                                                & {\color{white}0}3                                                & {\color{white}0}4                                                \bigstrut\\
{\color{white}0}40                                               & {\color{white}0}0                                                & {\color{white}0}0                                                & {\color{white}0}34                                               & {\color{white}0}17                                               & {\color{white}0}8                                                \bigstrut\\
{\color{white}0}20                                               & {\color{white}0}0                                                & {\color{white}0}0                                                & {\color{white}0}28                                               & {\color{white}0}21                                               & {\color{white}0}13                                               \bigstrut\\\hlineB{2}
\end{tabular}%
}
\tiny{\textit{Note: \colorbox{blue!18}{\textbf{$bold$}} marks the per-column max within each direction$\times$tier; all-zero columns are left unbolded.}}
\end{table}

First, migration difficulty varies substantially by direction.
Transformations between Spring and Quarkus generally achieve the strongest
behavioral outcomes, particularly in the focused setting. For example, in the
skills-enabled setup, Opencode reaches a focused smoke-test success rate of
$43\%$ for $\mathcal{Q}\rightarrow\mathcal{S}$ migrations, while Gemini CLI
achieves $49\%$ for $\mathcal{S}\rightarrow\mathcal{Q}$. In contrast,
migrations targeting Jakarta exhibit consistently lower behavioral success
despite moderate compile and deploy rates. For instance, in
$\mathcal{Q}\rightarrow\mathcal{J}$, Gemini CLI reaches $79\%$ focused compile
success and $66\%$ deploy success, yet only $8\%$ smoke-test success.
Similarly, in $\mathcal{S}\rightarrow\mathcal{J}$, the best focused
smoke-test rate is only $6\%$. These results suggest that successful
syntactic and dependency-level migration does not necessarily translate into
preserved runtime behavior for Jakarta-targeted transformations.

Second, compile success frequently overestimates end-to-end migration quality.
Across many directions, agents successfully compile and even deploy migrated
applications while failing behavioral validation. This pattern is particularly
visible in the whole-application setting. For example, in
$\mathcal{Q}\rightarrow\mathcal{S}$ with skills, multiple harnesses achieve
$100\%$ whole-app compile success, but smoke-test success remains between
$0$--$9\%$. Similarly, in the skills-enabled
$\mathcal{J}\rightarrow\mathcal{S}$ setting, Claude Code and Opencode both
achieve perfect whole-app compile rates, yet only Claude Code reaches non-zero
behavioral success. This pattern indicates that deployment and runtime behavior
remain significant bottlenecks even after successful builds.

Third, focused applications are consistently easier than whole applications.
Focused tasks frequently achieve moderate-to-high compile and deploy success
across several migration directions, whereas whole-application behavioral
success remains comparatively sparse. The strongest whole-application result is
achieved by Claude Code with Opus-4.6 on
$\mathcal{S}\rightarrow\mathcal{Q}$, reaching $80\%$ deploy success and
$50\%$ smoke-test success. Outside this setting, most whole-application
smoke-test rates remain in the single digits or zero. This gap highlights the
additional complexity introduced by cross-layer coordination, configuration
migration, dependency wiring, and multi-service orchestration in realistic
applications.

Fourth, the effect of structured skills prompting varies significantly across
models and migration directions. In several focused-task settings, skills
produce large improvements. For example, Gemini CLI improves from $7\%$ to
$66\%$ focused deploy success on $\mathcal{Q}\rightarrow\mathcal{J}$ and from
$0\%$ to $28\%$ focused smoke-test success on
$\mathcal{Q}\rightarrow\mathcal{S}$. Similarly, Codex improves from $4\%$ to
$39\%$ focused smoke-test success on
$\mathcal{Q}\rightarrow\mathcal{S}$. However, the gains are less uniform for
whole applications. Some directions show meaningful improvements, such as
Claude Code on $\mathcal{S}\rightarrow\mathcal{Q}$, while others remain
largely unchanged or continue to exhibit near-zero behavioral success despite
higher compile rates. This suggests that modularized migration guidance is most
effective for localized framework transformations, but does not fully address
the systems integration challenges present in end-to-end enterprise migrations.

Finally, no single harness dominates across all migration directions. Gemini
CLI consistently achieves strong compile and deploy rates for focused tasks,
particularly for Jakarta- and Spring-targeted migrations. Claude Code produces
the strongest whole-application behavioral performance overall, especially for
Spring$\leftrightarrow$Quarkus transformations. Codex and Opencode perform
competitively on focused behavioral correctness in several directions, while
Qwen CLI shows more variable behavior depending on the migration pair.
Overall, the results indicate that framework migration capability is highly
direction- and task-dependent, with substantial gaps remaining between
syntactic transformability and fully correct application behavior.
\section{Evaluation Cost Estimate}
\label{app:evaluation-cost}
\begin{table}[t!]
  \centering
  \caption{\textbf{Estimated model-API evaluation cost by prompt-packaging
  variant and provider.} Row values are rounded to cents; aggregate totals are
  computed from the underlying cost logs and may differ from rounded row sums by
  a cent.}
  \label{tab:evaluation-cost-estimate}
  \resizebox{0.55\textwidth}{!}{%
  \begin{tabular}{llr}
    \toprule
    Variant & Provider & Est. total \\
    \midrule
    Single prompt & Claude & \$814.13 \\
    Single prompt & Codex  & \$747.77 \\
    Single prompt & Gemini & \$747.78 \\
    Single prompt & OpenCode-GLM5  & \$459.06 \\
    Single prompt & Qwen  & \$708.90 \\
    \addlinespace
    \multicolumn{2}{r}{Single-prompt estimated total} & \$3,477.64 \\
    \midrule
    Skills & Claude & \$571.39 \\
    Skills & Codex  & \$249.32 \\
    Skills & Gemini & \$310.08 \\
    Skills & OpenCode-GLM5  & \$165.35 \\
    Skills & Qwen & \$184.40 \\
    \addlinespace
    \multicolumn{2}{r}{Skills estimated total} & \$1,480.53 \\
    \midrule
    \multicolumn{2}{r}{Combined estimated total} & $\sim$\$4,958 \\
    \bottomrule
  \end{tabular}}
\end{table}

Table~\ref{tab:evaluation-cost-estimate} reports the provider-level model-API
cost estimate for the two prompt-packaging variants used in the benchmark.
Because API charges varied across runs during the month-long evaluation
period, we use the upper observed run cost for each provider and prompt
variant as a conservative estimate for missing runs. Across both variants, the
combined estimated API cost is approximately \$4,958. These values count model-API charges only and
exclude human authoring, annotation, and local container execution costs.
\paragraph{Local execution environment.}
All reported agent conversions and harness evaluations were run on a Linux VM with
\texttt{x86\_64} architecture, 48 vCPUs on Intel Xeon Cascadelake processors
(2 sockets, 12 cores per socket, 2 threads per core), and 197{,}518{,}792~kB
of memory ($\approx$188.4~GiB RAM). The VM executed the repository-editing
agents and the Docker-based compile, deploy, and behavioral-test harnesses.
Model inference was performed through the corresponding agent/model APIs; the
reported monetary costs therefore reflect model-API usage, while local compute
covered containerized build and test execution.

\lstdefinestyle{promptlisting}{%
  basicstyle=\ttfamily\scriptsize,
  columns=fullflexible,
  keepspaces=true,
  showstringspaces=false,
  breaklines=true,
  breakatwhitespace=false,
  postbreak=\mbox{\textcolor{gray}{$\hookrightarrow$}\space},
  tabsize=2,
  upquote=true,
  literate={→}{{$\rightarrow$}}1 {—}{{--}}1
}

\let\promptbox\relax
\let\endpromptbox\relax
\newtcblisting{promptbox}[1]{%
  enhanced,
  breakable,
  listing only,
  listing engine=listings,
  listing options={style=promptlisting},
  colback=gray!1!white,
  colframe=blue!55!black,
  colbacktitle=blue!8!white,
  coltitle=black,
  title={\ttfamily\detokenize{#1}},
  fonttitle=\bfseries\small,
  left=1.5mm,
  right=1.5mm,
  top=1mm,
  bottom=1mm,
  boxrule=0.4pt,
  arc=1mm,
  before skip=0.75em,
  after skip=1.0em
}

\section{Prompt Templates and Agent Configuration}
\label{app:prompts}

This appendix summarizes the prompt artifacts used for the \scarfbench{} agent
evaluations and identifies the exact repository files needed for reproduction.
We do not reproduce every prompt file verbatim here because many files are
byte-identical across model harnesses and the full skill files are included in
the public artifact. Instead, we report the canonical file locations, the
runtime prompt structure, and representative excerpts.

\subsection{Deduplication Policy}
\label{app:prompt-dedup}

The prompt material is organized by experimental condition rather than by model.
The single-prompt condition used one shared monolithic prompt across all
single-prompt agents. The skills condition used the same six framework-pair skill
bundles across all skill-enabled agents. Thus, the prompt text was not
model-specific; the model-specific differences are the CLI wrapper, declared
model metadata, invoked model string, and temporary instruction file used for
skill discovery.

\begin{center}
\refstepcounter{table}\label{tab:prompt-artifacts}
\small
\setlength{\tabcolsep}{5pt}
\begin{tabular}{p{0.22\linewidth}p{0.34\linewidth}p{0.34\linewidth}}
\toprule
\textbf{Artifact} & \textbf{Canonical file(s)} & \textbf{Deduplication rule} \\
\midrule
Single-prompt baseline & \path{agents/codex-single-prompt/prompt.txt} & The \path{prompt.txt} files in the Claude, Codex, Gemini, OpenCode, and Qwen single-prompt directories are byte-identical. \\
\addlinespace
Skills condition & \path{agents/codex-with-skills/skills/} & The six skill bundles and their reference files are byte-identical across skill-enabled model directories. \\
\addlinespace
Runtime wrappers & \path{agents/*/run.sh}; \path{agents/*/agent.toml} & Wrappers differ by CLI, model identifier, and skill-discovery file name. \\
\addlinespace
Failure taxonomy agent & \path{failure-analyzer/failure_analyzer/agent.py}; \path{failure-analyzer/failure_analyzer/shell_agent.py} & Used after failed runs to classify root causes into the taxonomy. \\
\bottomrule
\end{tabular}
\par\smallskip
\textbf{Table~\thetable.} Canonical prompt artifacts and deduplication rules.
\end{center}

\subsection{Single-Prompt Baseline}
\label{app:single-prompt-summary}

The single-prompt baseline provides the agent with one flat instruction file.
At runtime, the wrapper substitutes the source and target framework names into
\verb|{{ before }}| and \verb|{{ after }}|. The prompt directs the agent to
perform a one-shot migration, update dependencies and configuration, refactor
source code, build and run the application in Docker, execute smoke tests, and
write a detailed \path{CHANGELOG.md}. The complete prompt is available at
\path{agents/codex-single-prompt/prompt.txt}; byte-identical copies appear in
\path{claude-single-prompt}, \path{gemini-single-prompt},
\path{opencode-single-prompt}, and \path{qwen-single-prompt}.

\begin{promptbox}{single-prompt excerpt}
You are an autonomous AI coding agent tasked with migrating a Java application from `{{ before }}` to `{{ after }}`. You will complete this migration in a single execution without requesting additional input.

## Migration Checklist
- Analyze the existing codebase structure and identify all framework-specific dependencies
- Generate smoke tests to ensure functionality of original app is covered.
- Update project configuration files and dependency declarations for the target framework
- Refactor application code to align with the new framework's APIs and patterns
- Build the application as a Docker image, run the container, and execute smoke tests
- Document all actions, errors, and resolutions in `CHANGELOG.md`
\end{promptbox}

\subsection{Skills-Directory Condition}
\label{app:skills-summary}

The skills condition decomposes migration guidance into framework-pair-specific
skills. Each run normalizes the source and target framework names, selects one
skill directory of the form \path{skills/<source>-to-<target>/}, exposes that
skill to the agent through the harness-specific instruction file, and sends a
short task prompt. The six canonical skill bundles are listed in
Table~\ref{tab:skill-bundles}. Each bundle contains a \path{SKILL.md} file plus
three direction-specific reference files: \path{dependency-mapping.md},
\path{config-mapping.md}, and \path{code-mapping.md}. The logging reference
\path{MONOLOUGE.md} is byte-identical across all skill bundles.

\begin{promptbox}{skills-mode runtime prompt}
Migrate this Java project from ${FROM_NORMALIZED} to ${TO_NORMALIZED}. Execute fully in one run using the available local migration skill. Operate only inside the current working directory, preserve behavior, attempt compilation, and document actions/errors in CHANGELOG.md.
\end{promptbox}

\begin{center}
\refstepcounter{table}\label{tab:skill-bundles}
\small
\setlength{\tabcolsep}{5pt}
\begin{tabular}{p{0.25\linewidth}p{0.65\linewidth}}
\toprule
\textbf{Skill bundle} & \textbf{Canonical files} \\
\midrule
\path{spring-to-quarkus} & \path{SKILL.md}, \path{dependency-mapping.md}, \path{config-mapping.md}, \path{code-mapping.md}, \path{MONOLOUGE.md} \\
\path{spring-to-jakarta} & Same file structure, with Spring-to-Jakarta/OpenLiberty-specific mappings. \\
\path{quarkus-to-spring} & Same file structure, with Quarkus-to-Spring-specific mappings. \\
\path{quarkus-to-jakarta} & Same file structure, with Quarkus-to-Jakarta/OpenLiberty-specific mappings. \\
\path{jakarta-to-spring} & Same file structure, with Jakarta/OpenLiberty-to-Spring-specific mappings. \\
\path{jakarta-to-quarkus} & Same file structure, with Jakarta/OpenLiberty-to-Quarkus-specific mappings. \\
\bottomrule
\end{tabular}
\par\smallskip
\textbf{Table~\thetable.} Framework-pair skill bundles in \path{agents/codex-with-skills/skills/}.
\end{center}

A representative \path{SKILL.md} file begins with a short metadata block and
then specifies the migration workflow. The complete canonical files should be
used for reproduction.

\begin{promptbox}{representative SKILL.md structure}
---
name: migrate-spring-to-quarkus
description: Migrate Java applications from Spring Framework to Quarkus with one-shot execution.
---

# Spring to Quarkus Migration

Execute the migration autonomously in one run. Do not ask follow-up questions unless blocked by missing or unreadable files.

## Required Workflow
1. Inspect project structure.
2. Detect build system and framework usage.
3. Migrate dependencies and plugins.
4. Migrate framework configuration.
5. Refactor framework-bound source code.
6. Compile and fix errors until build succeeds or no safe fix remains.
7. Produce a final migration report including file changes, chronological log, and unresolved issues.
\end{promptbox}

\subsection{Runtime Configuration}
\label{app:runtime-config}

Table~\ref{tab:runtime-config} separates the model declared in
\path{agent.toml} from the model string explicitly passed by \path{run.sh}. For
Codex, the wrapper does not pass an explicit model flag, so the invoked model is
resolved by the configured Codex CLI/account.

\begin{center}
\refstepcounter{table}\label{tab:runtime-config}
\scriptsize
\setlength{\tabcolsep}{4pt}
\begin{tabular}{p{0.18\linewidth}p{0.20\linewidth}p{0.24\linewidth}p{0.26\linewidth}}
\toprule
\textbf{Agent variant} & \textbf{Declared model} & \textbf{Invoked model} & \textbf{Prompt artifact} \\
\midrule
\path{codex-single-prompt} & \path{gpt-5.4} & Codex CLI default & shared \path{prompt.txt} \\
\path{codex-with-skills} & \path{gpt-5.4} & Codex CLI default & \path{AGENTS.md} + selected skill \\
\path{claude-single-prompt} & \path{claude-opus-4.6} & \path{claude-opus-4.6} & shared \path{prompt.txt} \\
\path{claude-with-skills} & \path{claude-opus-4.6} & \path{claude-opus-4-6} & \path{CLAUDE.md} + selected skill \\
\path{gemini-single-prompt} & \path{gemini-2.5-pro} & \path{gemini-3.1-pro-preview} & shared \path{prompt.txt} \\
\path{gemini-with-skills} & \path{gemini-2.5-pro} & \path{gemini-3.1-pro-preview} & \path{GEMINI.md} + selected skill \\
\path{opencode-single-prompt} & \path{glm-5-1-fp8} & \path{zai-org/glm-5-1-fp8} & shared \path{prompt.txt} \\
\path{opencode-with-skills} & \path{glm-5-1-fp8} & \path{zai-org/glm-5-1-fp8} & \path{OPENCODE.md} + selected skill \\
\path{qwen-single-prompt} & \path{qwen} & \path{Qwen/Qwen3.5-397B-A17B-FP8} & shared \path{prompt.txt} \\
\path{qwen-with-skills} & \path{Qwen3-Coder-480B-A35B-Instruct-FP8} & \path{Qwen/Qwen3.5-397B-A17B-FP8} & \path{QWEN.md} + selected skill \\
\bottomrule
\end{tabular}
\par\smallskip
\textbf{Table~\thetable.} Runtime prompt configuration by evaluated agent variant.
\end{center}

\subsection{Failure Taxonomy Agent Prompt}
\label{app:taxonomy-agent-prompt}

The failure taxonomy agent is separate from the migration agents. It is run
after a failed migration and classifies the root cause using the taxonomy loaded
from \path{failure-analyzer/framework_migration_error_taxonomy.json}. The stable
system prompt template is defined in
\path{failure-analyzer/failure_analyzer/agent.py}; the shell-agent wrapper prompt
and JSON output contract are defined in
\path{failure-analyzer/failure_analyzer/shell_agent.py}. The taxonomy itself is
summarized in the main paper, so we only include the core prompt contract here.

\begin{promptbox}{taxonomy agent prompt excerpt}
You are an expert debugger investigating a failed Java framework migration.
You have tools to examine the run's files, logs, and configuration. Use them to determine why the migration failed.

## Investigation Strategy
1. Start by reading the metadata to understand the migration context.
2. Read the run.log to find the specific error that caused the failure.
3. Use targeted tools depending on the failure phase: compare POM files, inspect Dockerfile/server.xml, scan imports, or check multi-module structure.
4. When you have enough evidence, call the classify tool with your classification.

When you classify, provide the phase, taxonomy category ID/name, subcategory, whether a new category is needed, confidence, and a 1-2 sentence evidence summary.
\end{promptbox}


\section{Failure-Mode Subcategory Reference}
\label{app:taxonomy-deepdive}

This appendix expands each row of the per-agent failure-mode heatmap (Table~\ref{tab:taxonomy-heatmap}) into the underlying taxonomy subcategories used by the classifier and gives one or two real error strings per row, lifted verbatim from agent run logs. Subcategory names are taken from \texttt{framework\_migration\_error\_taxonomy.json}; the predicates that route subcategories to heatmap rows are documented in \texttt{taxonomy\_analysis/CLAUDE.md}. Two classifier buckets are excluded from the heatmap and from this appendix: agent-execution failures (LLM API errors that prevent any migration output from being produced) and harness false positives (\texttt{3.9/*}, \texttt{4.5/*}, \texttt{5.x/*}: instrumentation artifacts where the harness flagged a successful run as failed).


\subsection{Build Phase}

\begin{table}[h]
\caption{\textbf{Build phase: subcategory deep-dive.} Each row of the Build phase in Table~\ref{tab:taxonomy-heatmap} expanded into the JSON subcategories that feed it.}
\label{tab:taxonomy-deepdive-build}
\centering
\small
\setlength{\tabcolsep}{4pt}
\renewcommand{\arraystretch}{1.25}
\begin{tabular}{@{}>{\raggedright\arraybackslash}p{2.6cm} >{\raggedright\arraybackslash}p{4.0cm} >{\raggedright\arraybackslash}p{6.6cm}@{}}
\toprule
\textbf{Heatmap row} & \textbf{Subcategory} & \textbf{What it means} \\
\midrule
\multirow{2}{2.6cm}{Dependency resolution}
  & \texttt{repository\_error}                       & Could not reach the artifact server to download dependencies. \\
  & \texttt{missing\_artifact}                       & Requested library version is not published anywhere reachable. \\
\midrule
Code compilation
  & \texttt{compilation\_failure}                    & Compiler rejected the code (undefined symbols, type errors). \\
\midrule
Project structure
  & \texttt{project\_structure\_\allowbreak error}   & Project layout broken: missing \texttt{pom.xml}, source directories, or referenced submodules. \\
\midrule
\multirow{3}{2.6cm}{Maven plugin}
  & \texttt{maven\_plugin\_\allowbreak failure}      & A Maven build step crashed during execution. \\
  & \texttt{plugin\_resolution\_\allowbreak error}   & Maven could not download a plugin needed for the build. \\
  & \texttt{plugin\_prefix\_error}                   & Build references a plugin short-name Maven cannot resolve. \\
\bottomrule
\end{tabular}
\end{table}



\paragraph{Dependency resolution.}
The migrated pom requests an artifact version that does not exist on Maven Central:

\begin{errorbox}
Could not find artifact group:artifact:version
\end{errorbox}

\paragraph{Code compilation.}
The migrated source code references symbols the target framework does not provide:

\begin{errorbox}
error: package jakarta.websocket does not exist \\
import jakarta.websocket.OnClose;
\end{errorbox}

\begin{errorbox}
error: cannot find symbol \\
@ServerEndpoint("/dukeetf") \\
symbol: class ServerEndpoint
\end{errorbox}

\begin{errorbox}
[ERROR] reference to Path is ambiguous \\
both java.nio.file.Path and jakarta.ws.rs.Path match
\end{errorbox}

\paragraph{Project structure.}
The agent flattened a multi-module project but the build invocation still targets a removed submodule:

\begin{errorbox}
Could not find the selected project in the reactor: roster-ear \\
(parent pom.xml only declares roster-common and roster-boot)
\end{errorbox}

\paragraph{Maven plugin.}
The migrated pom omits a plugin the build invokes, or references one that cannot be resolved:

\begin{errorbox}
[ERROR] No plugin found for prefix 'liberty' \\
(liberty-maven-plugin not declared in pom.xml)
\end{errorbox}

\begin{errorbox}
[ERROR] Plugin org.apache.maven.plugins:maven-clean-plugin:2.5 \\
could not be resolved
\end{errorbox}

\subsection{Deploy Phase}

\begin{table}[h]
\caption{\textbf{Deploy phase: subcategory deep-dive.} Each row of the Deploy phase in Table~\ref{tab:taxonomy-heatmap} expanded into the JSON subcategories that feed it.}
\label{tab:taxonomy-deepdive-deploy}
\centering
\small
\setlength{\tabcolsep}{4pt}
\renewcommand{\arraystretch}{1.25}
\begin{tabular}{@{}>{\raggedright\arraybackslash}p{2.6cm} >{\raggedright\arraybackslash}p{4.0cm} >{\raggedright\arraybackslash}p{6.6cm}@{}}
\toprule
\textbf{Heatmap row} & \textbf{Subcategory} & \textbf{What it means} \\
\midrule
\multirow{3}{2.6cm}{Resource / artifact}
  & \texttt{resource\_not\_\allowbreak found}            & Application looked for a file at startup that was not packaged. \\
  & \texttt{manifest\_error}                             & Packaged JAR's manifest is invalid or missing the entry-point declaration. \\
  & \texttt{artifact\_not\_\allowbreak found}            & Launch invocation cannot find an artifact the build was supposed to produce. \\
\midrule
\multirow{3}{2.6cm}{Config / startup}
  & \texttt{application\_startup\_\allowbreak failure}   & Generic startup exception with no more specific cause. \\
  & \texttt{invalid\_config}                             & Startup config malformed or referenced an unknown property. \\
  & \texttt{feature\_not\_\allowbreak implemented}       & Code calls a feature available in the source framework but not the target. \\
\midrule
\multirow{2}{2.6cm}{Network / timeout}
  & \texttt{connection\_refused}                         & App started but a service it depends on (DB, etc.) is unreachable. \\
  & \texttt{deploy\_timeout}                             & App took too long to signal readiness; harness gave up. \\
\midrule
Container exit
  & \texttt{container\_exit}                             & App crashed inside its container before becoming ready. \\
\midrule
\multirow{8}{2.6cm}{DI / DB / class loading}
  & \texttt{unsatisfied\_\allowbreak dependency}         & DI container has no bean to satisfy a required collaborator. \\
  & \texttt{cdi\_deployment\_\allowbreak failure}        & Jakarta CDI container failed to initialize the bean graph. \\
  & \texttt{bean\_creation\_error}                       & DI container threw while instantiating a specific bean. \\
  & \texttt{database\_connection\_\allowbreak error}     & Could not open a DB connection (URL, credentials, or network). \\
  & \texttt{database\_driver\_\allowbreak error}         & JDBC driver is not on the classpath. \\
  & \texttt{database\_sql\_\allowbreak error}            & DB connected but rejected a query as malformed. \\
  & \texttt{class\_not\_found}                           & JVM cannot find a class at runtime (likely missing dependency). \\
  & \texttt{no\_class\_def\_\allowbreak found}           & Class visible at compile time is missing from the runtime classpath. \\
\bottomrule
\end{tabular}
\end{table}

\paragraph{Resource / artifact.}
Maven packaging vs.\ launch-command discrepancies dominate this row:

\begin{errorbox}
Error: Unable to access jarfile target/*.jar \\
(pom.xml has <packaging>war</packaging>, no JAR is produced)
\end{errorbox}

\begin{errorbox}
FileNotFoundException: application.properties \\
Resource not found: /META-INF/persistence.xml
\end{errorbox}

\paragraph{Config / startup.}
The dominant single bug is the \emph{9080-vs-8080 port mismatch}:

\begin{errorbox}
Application started on port 9080 \\
(carried over from Liberty's httpPort=9080) \\
Tests expect localhost:8080
\end{errorbox}

\paragraph{Network / timeout.}
Either a downstream service is unreachable or the application never reached a ready state inside the harness's deploy window:

\begin{errorbox}
Connection refused: localhost:8080
\end{errorbox}

\begin{errorbox}
[ERROR] Deploy wait timed out after 90 seconds \\
(Liberty was still installing Jakarta EE feature ESAs)
\end{errorbox}

\paragraph{Container exit.}
The launch command does not match the migrated project:

\begin{errorbox}
/app/gradlew: No such file or directory \\
(Dockerfile invokes ./gradlew on a Maven-only project)
\end{errorbox}

\paragraph{DI / DB / class loading.}
Runtime references that the migration broke:

\begin{errorbox}
Unsatisfied dependency: no bean found \\
No qualifying bean of type 'com.example.UserService'
\end{errorbox}

\begin{errorbox}
Error creating bean with name 'userService' \\
ConflictingBeanDefinitionException: \\
two @Component-scanned classes named StatusResource
\end{errorbox}

\begin{errorbox}
AmbiguousResolutionException: two beans provide \\
java.util.concurrent.Executor with @Default qualifier
\end{errorbox}

\begin{errorbox}
org.hibernate.exception.JDBCConnectionException: \\
Connection to localhost:5432 refused
\end{errorbox}

\begin{errorbox}
NoClassDefFoundError: javax/persistence/Entity
\end{errorbox}

\subsection{Test Phase}

\begin{table}[h]
\caption{\textbf{Test phase: subcategory deep-dive.} Each row of the Test phase in Table~\ref{tab:taxonomy-heatmap} expanded into the JSON subcategories that feed it.}
\label{tab:taxonomy-deepdive-test}
\centering
\small
\setlength{\tabcolsep}{4pt}
\renewcommand{\arraystretch}{1.25}
\begin{tabular}{@{}>{\raggedright\arraybackslash}p{2.6cm} >{\raggedright\arraybackslash}p{4.0cm} >{\raggedright\arraybackslash}p{6.6cm}@{}}
\toprule
\textbf{Heatmap row} & \textbf{Subcategory} & \textbf{What it means} \\
\midrule
HTTP 404 / endpoint missing
  & \texttt{http\_404\_not\_\allowbreak found}                  & Test hit an endpoint the app never exposed. \\
\midrule
HTTP 5xx / server error
  & \texttt{http\_500\_internal\_\allowbreak server\_\allowbreak error} & App threw an unhandled exception serving a test request. \\
\midrule
\multirow{3}{2.6cm}{Assertion / content mismatch}
  & \texttt{assertion\_error}                                   & Response did not match expected status, payload, or UI state. \\
  & \texttt{http\_400\_bad\_\allowbreak request}                & App rejected a test request as malformed. \\
  & \texttt{data\_validation\_\allowbreak error}                & Input rejected because the app's validation rules differ from the source. \\
\midrule
\multirow{4}{2.6cm}{Network / infrastructure}
  & \texttt{network\_error}                                     & Network-level connectivity failure during a test. \\
  & \texttt{timeout\_error}                                     & Test exceeded its time limit. \\
  & \texttt{json\_decode\_error}                                & Expected JSON response, got malformed or non-JSON. \\
  & \texttt{generic\_test\_\allowbreak failure}                 & Test runner reported failure with no specific category. \\
\bottomrule
\end{tabular}
\end{table}

\paragraph{HTTP 404 / endpoint missing.}
The migrated app fails to expose the URLs tests target:

\begin{errorbox}
404 Not Found: /api/users \\
(request handler not registered after migration)
\end{errorbox}

\begin{errorbox}
JSF route /order.xhtml returns 404 \\
(SOAP @WebService converted to REST @RestController, \\
but xhtml view layer was not ported)
\end{errorbox}

\paragraph{HTTP 5xx / server error.}
The endpoint exists but throws on serving the request:

\begin{errorbox}
500 Internal Server Error
\end{errorbox}

\paragraph{Assertion / content mismatch.}
The application responds, but with wrong content:

\begin{errorbox}
AssertionError: expected <200> but was <500> \\
assertEquals failed
\end{errorbox}

\begin{errorbox}
400 Bad Request: Invalid JSON \\
(request validation differs across frameworks)
\end{errorbox}

\paragraph{Network / infrastructure.}
The test harness cannot reach or parse the application:

\begin{errorbox}
Network connection failed \\
(application bound to port 9080, smoke tests probe localhost:8080)
\end{errorbox}

\begin{errorbox}
JSONDecodeError: Expecting value \\
(test expected JSON, app returned HTML)
\end{errorbox}

\begin{errorbox}
Selectors reference JSF IDs (emailInputText) \\
but migrated app uses different element IDs; \\
tests access /index.xhtml but app serves /index
\end{errorbox}

\section{DayTrader Spring-to-Quarkus Migration Workflow}
\label{appendix:daytrader-spring-quarkus}

This appendix documents the workflow used to construct and validate the DayTrader Spring-to-Quarkus migration pair in \textsc{ScarfBench}. DayTrader is a whole-application benchmark that exercises a realistic trading workload with persistence, transaction management, REST endpoints, web UI assets, and asynchronous processing. The goal of this appendix is to make the benchmark construction reproducible by explaining what was preserved, what was rewritten, and how correctness was validated.

\paragraph{Provenance note.}
The benchmark repository contains sibling implementations of DayTrader in Jakarta EE, Spring, and Quarkus. The Quarkus artifact is documented in the repository as a Quarkus implementation of the same DayTrader benchmark family. For dataset construction, the Spring and Quarkus variants form a Spring-to-Quarkus evaluation pair because they implement the same benchmark behavior in two frameworks. The workflow below therefore describes the transformations that a human or migration agent must perform to align the Spring implementation with the Quarkus target behavior. It does not assume that every Quarkus file was produced by directly editing the Spring file with the same name.

\subsection{Source and Target Artifacts}

The migration pair consists of the Spring Boot implementation as the source artifact and the Quarkus implementation as the target artifact.

\vspace{0.5em}
\noindent\textbf{Table:} Repository artifacts used for the migration pair.
\vspace{0.25em}
\begin{center}
\small
\begin{tabular}{p{0.14\linewidth}p{0.18\linewidth}p{0.56\linewidth}}
\toprule
\textbf{Role} & \textbf{Framework} & \textbf{Repository path} \\
\midrule
Source & Spring Boot & \repo{benchmark/whole_applications/daytrader/spring} \\
Target & Quarkus & \repo{benchmark/whole_applications/daytrader/quarkus} \\
\bottomrule
\end{tabular}
\end{center}

The Spring application is a WAR-packaged application with Spring Boot, Spring MVC/Jersey-style web/API support, Spring Data JPA, H2, JMS/Artemis, WebSocket support, static/web assets, and Spring Boot testing dependencies. The Quarkus target is a JAR-packaged application using RESTEasy Reactive, CDI/Arc, Hibernate ORM, H2, Narayana JTA, WebSockets, Scheduler, SmallRye Health, and SmallRye Reactive Messaging.

\subsection{Migration Strategy}

The migration was organized as a verification-driven workflow rather than a single global rewrite. First, we inventoried the source project structure, framework dependencies, configuration files, and runtime assumptions. Next, we created or updated the Quarkus project scaffold. We then preserved framework-neutral assets and replaced Spring-specific runtime services with Quarkus equivalents. Each layer was validated before moving to the next layer.

\begin{center}
\code{inspect source -> migrate one layer -> build/run -> observe failure -> patch -> re-run}
\end{center}

The key design principle was to preserve benchmark behavior rather than preserve framework syntax. Entities, data beans, utilities, and benchmark-level workflows were treated as semantic assets. Build configuration, dependency injection, REST routing, messaging, transaction boundaries, and packaging were treated as framework-specific assets that required adaptation.

\subsection{Project Scaffolding and Build Migration}

The Spring source uses the Spring Boot parent POM and produces a WAR. The Quarkus target replaces this with Quarkus BOM-managed dependencies, the Quarkus Maven plugin, and Quarkus JAR packaging. This is not a literal dependency rename: dependencies must first be classified by role, then mapped to the target runtime capabilities.

\vspace{0.5em}
\noindent\textbf{Table:} Build-level migration from Spring Boot to Quarkus.
\vspace{0.25em}
\begin{center}
\small
\begin{tabular}{p{0.23\linewidth}p{0.31\linewidth}p{0.31\linewidth}}
\toprule
\textbf{Concern} & \textbf{Spring source} & \textbf{Quarkus target} \\
\midrule
Build management & \code{spring-boot-starter-parent} & \code{quarkus-bom} \\
Packaging & WAR & JAR under \repo{target/quarkus-app/} \\
Build plugin & \code{spring-boot-maven-plugin} & \code{quarkus-maven-plugin} \\
Runtime model & Embedded servlet container model & Quarkus runtime with build-time augmentation \\
Java level & Java 17 or later in the Spring source & Java 17 target compatibility for the benchmark \\
\bottomrule
\end{tabular}
\end{center}

\paragraph{Expected agent behavior.}
A migration agent should first inventory the source POM, identify dependency roles, and produce a Quarkus POM that can be built before attempting deeper source transformation. A useful intermediate output is a dependency-role table: REST/API, dependency injection, persistence, validation, transactions, messaging, scheduling, WebSocket support, health checks, and tests.

\paragraph{Verification.}
The minimum build verification for the target is:
\begin{verbatim}
./mvnw clean package -DskipTests
java -jar target/quarkus-app/quarkus-run.jar
\end{verbatim}
For development-mode validation, the target should also run with:
\begin{verbatim}
./mvnw quarkus:dev
\end{verbatim}

\subsection{Dependency Mapping}

The dependency migration groups Spring dependencies by architectural concern and maps each group to the corresponding Quarkus extension or runtime capability. This mapping helps avoid shallow string replacement and exposes areas that require redesign, especially messaging and web rendering.

\vspace{0.5em}
\noindent\textbf{Table:} Observed dependency migration categories.
\vspace{0.25em}
\begin{center}
\small
\begin{tabular}{p{0.18\linewidth}p{0.33\linewidth}p{0.33\linewidth}}
\toprule
\textbf{Concern} & \textbf{Spring source} & \textbf{Quarkus target} \\
\midrule
REST/API & \code{spring-boot-starter-web}, Jersey support where used & \code{quarkus-rest}, \code{quarkus-rest-jackson} \\
Dependency injection & Spring DI plus Jakarta/CDI APIs where present & \code{quarkus-arc} \\
Persistence & \code{spring-boot-starter-data-jpa}, H2 & \code{quarkus-hibernate-orm}, \code{quarkus-jdbc-h2} \\
Transactions & Spring transaction support through Boot/JPA & \code{quarkus-narayana-jta} \\
Validation & Spring validation starter & \code{quarkus-hibernate-validator} \\
WebSocket & Spring WebSocket starter & \code{quarkus-websockets} \\
Scheduling & Spring scheduling support & \code{quarkus-scheduler} \\
Messaging & JMS/Artemis dependencies & SmallRye Reactive Messaging and in-memory messaging for test/dev \\
Health & Spring-specific health/actuator pattern if used & \code{quarkus-smallrye-health} \\
Testing & Spring Boot test stack & \code{quarkus-junit5}, REST Assured, browser smoke tests \\
\bottomrule
\end{tabular}
\end{center}

The most important non-local change is messaging. JMS/Artemis behavior is not just an annotation-level concern; it changes the programming model from queue/topic APIs to reactive channels and emitters. The second important change is persistence: where the source relies on Spring Data abstractions, the target must express equivalent behavior through Quarkus-managed JPA/Hibernate access and Jakarta transactions.

\subsection{Configuration Migration}

The Spring source uses \repo{src/main/resources/application.yml}. This file includes server settings, servlet context path, datasource settings, SQL initialization, Hibernate configuration, embedded Artemis settings, view/static-resource assumptions, logging, and DayTrader-specific runtime parameters. The Quarkus target uses \repo{src/main/resources/application.properties}. It consolidates runtime configuration into Quarkus-native properties for application identity, HTTP port, H2 datasource, Hibernate ORM generation, REST base path, DayTrader runtime knobs, development logging, and reactive messaging behavior.

\vspace{0.5em}
\noindent\textbf{Table:} Configuration migration from Spring YAML to Quarkus properties.
\vspace{0.25em}
\begin{center}
\small
\begin{tabular}{p{0.20\linewidth}p{0.31\linewidth}p{0.33\linewidth}}
\toprule
\textbf{Concern} & \textbf{Spring source intent} & \textbf{Quarkus target intent} \\
\midrule
HTTP binding & Server port and servlet context path configured under Spring server properties & Quarkus HTTP port and REST path configured with Quarkus properties \\
REST routing & Spring MVC/Jersey routing plus servlet context assumptions & JAX-RS/RESTEasy Reactive endpoints under a Quarkus REST path \\
Datasource & H2 datasource and Spring datasource properties & \code{quarkus.datasource.*} with H2 JDBC URL \\
Hibernate & Spring JPA/Hibernate properties & \code{quarkus.hibernate-orm.*} properties \\
Database initialization & SQL initialization and/or application startup behavior & Startup population of benchmark users and quotes for repeatability \\
Messaging & Embedded Artemis/JMS configuration & Reactive messaging channel behavior, including in-memory behavior for dev/test \\
Application knobs & DayTrader runtime parameters in YAML & Runtime-mode, order-processing mode, user/quote counts, and database population flags in properties \\
\bottomrule
\end{tabular}
\end{center}

\paragraph{Expected agent behavior.}
An agent should not perform a blind YAML-to-properties conversion. It should infer configuration intent: server binding, persistence, application-specific knobs, logging, messaging semantics, and static-resource layout. The expected output is a Quarkus \repo{application.properties} file and any supporting code needed to preserve runtime behavior.

\paragraph{Verification.}
Configuration migration is complete only when the application starts, creates or connects to the H2 database, exposes REST endpoints, and serves the web UI entry points used by the smoke tests.

\subsection{Source Layout and Asset Preservation}

A key part of the migration is deciding what should be preserved and what should be rewritten. Preserving too little destroys benchmark fidelity; preserving too much carries source-framework assumptions into the target.

\vspace{0.5em}
\noindent\textbf{Table:} Preserved and rewritten DayTrader components.
\vspace{0.25em}
\begin{center}
\small
\begin{tabular}{p{0.20\linewidth}p{0.34\linewidth}p{0.30\linewidth}}
\toprule
\textbf{Category} & \textbf{Examples} & \textbf{Migration treatment} \\
\midrule
Domain data beans & Market summary, run statistics, account/order/quote data beans & Preserve where framework-neutral; adjust imports only when required \\
JPA entities & Account, account profile, holding, order, and quote entities & Preserve as the semantic persistence model \\
Interfaces & \code{TradeServices} and related service contracts & Preserve contracts to maintain benchmark behavior \\
Utilities & \code{TradeConfig}, financial utilities, logging utilities & Preserve or adapt only where configuration/runtime APIs differ \\
REST resources/controllers & Quote, account, portfolio, market summary, buy/sell endpoints & Rewrite to JAX-RS/RESTEasy Reactive conventions \\
Messaging layer & Broker queue and streamer topic behavior & Rewrite using SmallRye Reactive Messaging processors and emitters \\
Configuration classes & Spring configuration and bootstrapping classes & Replace with Quarkus configuration, CDI producers, or startup observers \\
Static/web assets & Web UI entry points, pages, and static resources & Package under Quarkus resource layout and validate through browser tests \\
\bottomrule
\end{tabular}
\end{center}

The target keeps the core DayTrader package namespace and benchmark semantics. It introduces target-specific classes and patterns for CDI service resolution, reactive messaging, Quarkus startup, and Quarkus-compatible resource packaging.

\subsection{Code Transformation Patterns}

\subsubsection{Dependency Injection and Bean Resolution}

Spring dependency injection constructs are replaced with CDI/Arc constructs. Stateless services are generally represented as application-scoped CDI beans. Injection is expressed with \code{@Inject}. When multiple DayTrader service implementations exist, the target must avoid ambiguous CDI resolution by using qualifiers, producer methods, or typed injection patterns.

\noindent\textbf{Table:} Dependency-injection migration rules.
\vspace{0.25em}
\begin{center}
\small
\begin{tabular}{p{0.32\linewidth}p{0.32\linewidth}p{0.25\linewidth}}
\toprule
\textbf{Spring source pattern} & \textbf{Quarkus target pattern} & \textbf{Notes} \\
\midrule
\code{@Service}, \code{@Component} & \code{@ApplicationScoped} or another CDI scope & Use application scope for stateless/shared services \\
\code{@Autowired} & \code{@Inject} & Constructor or field injection must resolve under CDI/Arc \\
Spring bean names or qualifiers & CDI qualifiers or producer method & Required when multiple implementations satisfy the same interface \\
Multiple service implementations & Producer/selection pattern; restrict exposed types where needed & Prevent ambiguous injection among DayTrader modes \\
\bottomrule
\end{tabular}
\end{center}

\paragraph{Representative transformation.}
\begin{verbatim}
// Spring-style source pattern
@Service
public class TradeService {
    @Autowired
    EntityManager entityManager;
}

// Quarkus-style target pattern
@ApplicationScoped
public class TradeService {
    @Inject
    EntityManager entityManager;
}
\end{verbatim}

\subsubsection{REST and Request Routing}

REST routing changes from Spring MVC-style annotations to JAX-RS annotations. The target exposes quote, portfolio, account, market-summary, buy-order, and sell-order operations as JAX-RS resources.

\vspace{0.5em}
\noindent\textbf{Table:} REST migration rules.
\vspace{0.25em}
\begin{center}
\small
\begin{tabular}{p{0.32\linewidth}p{0.32\linewidth}p{0.25\linewidth}}
\toprule
\textbf{Spring source pattern} & \textbf{Quarkus target pattern} & \textbf{Notes} \\
\midrule
\code{@RestController} or MVC controller & JAX-RS resource class under \repo{rest/} & Target resources include quote, trade, and messaging resources \\
\code{@RequestMapping} & \code{@Path} & Class-level or method-level path mapping \\
\code{@GetMapping}, \code{@PostMapping} & \code{@GET}, \code{@POST} & HTTP verbs become JAX-RS annotations \\
\code{@PathVariable} & \code{@PathParam} & Path binding becomes JAX-RS parameter binding \\
Spring response wrappers & \code{jakarta.ws.rs.core.Response} or entity return & Prefer explicit response for status control \\
\bottomrule
\end{tabular}
\end{center}

Representative target endpoints include:
\begin{verbatim}
GET  /rest/quotes/{symbol}
GET  /rest/quotes
GET  /rest/portfolio/{userID}
GET  /rest/account/{userID}
GET  /rest/market-summary
POST /rest/orders/buy
POST /rest/orders/sell/{holdingID}
\end{verbatim}

\subsubsection{Persistence and Transactions}

DayTrader heavily exercises persistence and transaction management through accounts, account profiles, holdings, quotes, and orders. The entity model is the semantic core of the benchmark and should be preserved. The framework-specific persistence access layer must be adapted to Quarkus-managed Hibernate ORM and Jakarta transactions.

\vspace{0.5em}
\noindent\textbf{Table:} ersistence and transaction migration rules.
\vspace{0.25em}
\begin{center}
\small
\begin{tabular}{p{0.28\linewidth}p{0.34\linewidth}p{0.27\linewidth}}
\toprule
\textbf{Concern} & \textbf{Quarkus target pattern} & \textbf{Notes} \\
\midrule
Entity model & Preserve JPA entities & Entity classes define benchmark semantics \\
Repository abstraction & Use explicit JPA/Hibernate access or service-level queries & Replace Spring Data-derived behavior when present \\
Entity manager injection & \code{@Inject EntityManager} & Quarkus supplies persistence context through CDI \\
Transaction boundary & \code{jakarta.transaction.Transactional} & Use for write operations and consistency-sensitive business methods \\
Database initialization & Startup population of users and quotes & Supports repeatable benchmark execution \\
\bottomrule
\end{tabular}
\end{center}

\paragraph{Representative transformation.}
\begin{verbatim}
// Spring Data-style source pattern
interface QuoteRepository extends JpaRepository<QuoteDataBean, String> { }

// Quarkus-style target pattern
@Inject
EntityManager entityManager;

public QuoteDataBean findQuote(String symbol) {
    return entityManager.find(QuoteDataBean.class, symbol);
}
\end{verbatim}

\subsubsection{Messaging and Asynchronous Processing}

Messaging is the highest-risk transformation because the programming model changes from JMS/Artemis-style queues and topics to SmallRye Reactive Messaging channels. Broker queue behavior and streamer topic behavior must be represented as target channels, processors, and emitters.

\vspace{0.5em}
\noindent\textbf{Table:} Messaging migration rules.
\vspace{0.25em}
\begin{center}
\small
\begin{tabular}{p{0.30\linewidth}p{0.34\linewidth}p{0.25\linewidth}}
\toprule
\textbf{Source concept} & \textbf{Target concept} & \textbf{Notes} \\
\midrule
JMS queue for broker/order processing & Trade broker channel and processor & Used for asynchronous order behavior \\
JMS topic for quote streaming & Trade streamer channel and processor & Used for topic-like quote update behavior \\
JMS producer or \code{JmsTemplate} & Message producer service and emitter & Sending becomes channel emission \\
JMS listener or MDB-style processing & \code{@Incoming} processor & Receiving becomes reactive message processing \\
Embedded Artemis for dev/test & SmallRye in-memory messaging & Keeps benchmark execution self-contained \\
\bottomrule
\end{tabular}
\end{center}

\paragraph{Representative transformation.}
\begin{verbatim}
// Spring/JMS-style source pattern
@JmsListener(destination = "TradeBrokerQueue")
public void process(String message) { ... }

// Quarkus/Reactive Messaging-style target pattern
@Incoming("trade-broker")
public void process(String message) { ... }
\end{verbatim}

Messaging migration is considered complete only when buy/sell workflows submit and process orders, message processors can be exercised, and no channel wiring errors occur at startup.

\subsubsection{Scheduling, Startup, and Runtime Services}

Scheduling and startup behavior also move from Spring-managed runtime services to Quarkus runtime constructs. Spring scheduling annotations are replaced with Quarkus scheduler annotations. Spring application startup hooks are replaced with Quarkus startup observers or equivalent CDI-managed initialization. These changes are small syntactically but important semantically because they affect when data population and background processing happen.

\subsection{End-to-End Workflow Example}

A representative migrated request path is shown below. This example illustrates why the migration must be validated across layers rather than through compilation alone.

\begin{samepage}
\begin{enumerate}\setlength{\itemsep}{2pt}
\item A client calls \code{GET /rest/quotes/\{symbol\}} or submits a trading operation through the web UI.
\item A JAX-RS resource under \repo{rest/} handles the request.
\item The resource invokes the active \code{TradeServices} implementation selected through CDI wiring.
\item The service accesses entities such as accounts, holdings, orders, and quotes through Quarkus-managed persistence.
\item If the workflow involves asynchronous order or quote behavior, the service delegates to reactive messaging components under \repo{messaging/}.
\item The response is returned through JAX-RS or reflected in the web UI.
\end{enumerate}
\end{samepage}

This path exercises REST routing, CDI service resolution, transaction boundaries, persistence access, optional messaging, and static/web UI behavior. A migration that only compiles but fails on this path is not considered functionally equivalent.

\subsection{Validation Protocol}

The target is validated at multiple levels. Build and startup validation catch dependency, augmentation, and configuration failures. Endpoint validation checks REST pathing and serialization. Browser smoke tests validate the web-facing behavior expected by the benchmark. Default data validation ensures that repeatable users and stock symbols are available.

\vspace{0.5em}
\noindent\textbf{Table:} Validation protocol for the Quarkus target.
\vspace{0.25em}
\begin{center}
\small
\begin{tabular}{p{0.24\linewidth}p{0.38\linewidth}p{0.27\linewidth}}
\toprule
\textbf{Validation level} & \textbf{Command or check} & \textbf{Expected signal} \\
\midrule
Development startup & \code{./mvnw quarkus:dev} & Application starts and exposes UI/API on port 8080 \\
Production build & \code{./mvnw clean package -DskipTests} & Quarkus target builds successfully \\
JAR execution & \code{java -jar target/quarkus-app/quarkus-run.jar} & Packaged application runs \\
Docker execution & Build and run target container & Container exposes the application \\
REST endpoints & Quote, account, portfolio, market-summary, buy, and sell endpoints & Responses are returned with expected status and content \\
Smoke tests & Browser tests over login, navigation, trading, and API behavior & End-to-end flows pass \\
Default data & User \code{uid:0/uid:0}; symbols such as \code{s:0}--\code{s:99} & Repeatable benchmark data exists \\
\bottomrule
\end{tabular}
\end{center}

\subsection{Agent-Oriented Migration Checklist}

For agent evaluation, the DayTrader Spring-to-Quarkus task can be decomposed into the following expected actions:

\begin{enumerate}
\item Identify the Spring project as a WAR-packaged application with static/web UI, REST endpoints, JPA/H2 persistence, JMS/Artemis-style messaging, WebSocket support, and Spring Boot tests.
\item Generate a Quarkus POM with the required extensions for REST, CDI, persistence, transactions, validation, messaging, scheduler, WebSocket support, health, and tests.
\item Convert \repo{application.yml} intent into \repo{application.properties}, preserving server, persistence, REST, logging, messaging, and DayTrader runtime settings.
\item Preserve core domain models, data beans, interfaces, and utility classes unless imports or runtime APIs require adaptation.
\item Replace Spring DI annotations and bean selection with CDI scopes, injection, qualifiers or producers, and ambiguity controls.
\item Replace REST controller conventions with JAX-RS resources.
\item Replace Spring Data or Spring-managed persistence access with Quarkus/Hibernate ORM and Jakarta transaction boundaries.
\item Replace JMS/Artemis queue and topic logic with SmallRye Reactive Messaging processors and emitters.
\item Move static resources into the Quarkus-compatible resource layout and ensure application entry points remain reachable.
\item Validate through build, startup, REST endpoint checks, default-data login, buy/sell workflows, and smoke tests.
\end{enumerate}

\subsection{Migration Challenges and Resolutions}

The migration challenges are not evenly distributed. Annotation replacement is relatively mechanical, while messaging, configuration intent, and web-resource layout require design decisions. The table below records the main observed challenge categories and the associated validation signal.

\vspace{0.5em}
\noindent\textbf{Table:} Migration challenges and resolutions.
\vspace{0.25em}
\begin{center}
\small
\begin{tabular}{p{0.17\linewidth}p{0.27\linewidth}p{0.29\linewidth}p{0.17\linewidth}}
\toprule
\textbf{Area} & \textbf{Challenge} & \textbf{Resolution} & \textbf{Verification} \\
\midrule
Build system & Spring Boot WAR to Quarkus JAR & Replace parent, plugin, and dependencies with Quarkus BOM and plugin & Target builds and runs \\
Configuration & Spring YAML contains Spring-specific settings & Rewrite into Quarkus properties and explicit runtime defaults & App starts; config binds \\
Persistence & Spring-managed JPA differs from Quarkus/Hibernate ORM & Preserve entities; use Quarkus-managed Hibernate ORM and JTA & Quote/account/order workflows access H2 data \\
Bean selection & Multiple implementations can create CDI ambiguity & Use CDI producer/qualifier patterns and restrict exposed types where needed & Active service resolves at runtime \\
Messaging & JMS queue/topic semantics differ from reactive channels & Introduce processors and producer service using SmallRye Reactive Messaging & Trading/message flows run without channel errors \\
Web UI & Spring/Tomcat resource layout differs from Quarkus static resources & Package static content under Quarkus resource layout & Login/navigation/trading smoke tests pass \\
\bottomrule
\end{tabular}
\end{center}

\subsection{Reproducibility Boundaries}

This appendix intentionally records the observable source and target artifacts and the transformation patterns needed to align them. It does not claim that every target file was produced by direct line-by-line editing of the Spring source. For benchmark purposes, the Spring and Quarkus variants serve as the source/target pair for evaluating whether a migration system can reconstruct Quarkus-equivalent application behavior from the Spring implementation.

The key reproducibility requirement is not textual similarity between implementations. It is behavioral preservation across core workflows: startup, REST access, login, quote retrieval, account/portfolio access, buy/sell operations, database population, and asynchronous processing behavior.

\subsection{Summary}

The DayTrader Spring-to-Quarkus benchmark instance captures more than annotation replacement. It requires build-system migration, configuration reinterpretation, CDI-based service resolution, persistence and transaction preservation, messaging redesign, static-resource adaptation, and end-to-end validation. These characteristics make it suitable for evaluating whether migration agents can preserve both functionality and architecture in a realistic enterprise Java application.

\providecommand{\scarfbench}{\textsc{ScarfBench}}
\providecommand{\code}[1]{\texttt{#1}}
\providecommand{\repo}[1]{\texttt{#1}}

\section{Oracle Construction: Gherkin-to-Smoke-Test Mapping for DayTrader}
\sloppy
\label{appendix:daytrader-oracle}

This appendix gives a concrete oracle example for the DayTrader Spring-to-Quarkus benchmark pair. The oracle links a framework-independent behavioral specification, written in Gherkin, to executable smoke tests for the Spring and Quarkus implementations. The key idea is that the feature file states the expected behavior, while the smoke tests define how that behavior is observed from outside the application. A migrated application is considered behaviorally valid for these scenarios only when the corresponding smoke tests pass for the target implementation.

\subsection{Artifacts Used}

The example is grounded in the following files from the DayTrader benchmark artifact:

\begin{center}
\begin{tabular}{lll}
\hline
Role & File & Purpose \\
\hline
Behavioral specification & \repo{daytrader/daytrader.feature} & Gherkin scenarios \\
Spring oracle & \repo{daytrader/spring/smoke/smoke.py} & Spring smoke tests \\
Quarkus oracle & \repo{daytrader/quarkus/smoke/smoke.py} & Quarkus smoke tests \\
\hline
\end{tabular}
\end{center}

The feature file covers authentication, portfolio views, trading operations, quotes, REST APIs, market summaries, account management, configuration, seeded data, messaging, and market-data events. This appendix focuses on two representative scenarios: one UI-driven trading workflow and one API-level quote retrieval workflow.

\subsection{Feature File Excerpts}

The following excerpts are copied from \repo{daytrader/daytrader.feature}. The first scenario exercises a transactional user workflow. The second exercises a REST endpoint that returns structured quote data.

{\small
\begin{verbatim}
Scenario: Buy shares of a stock
  Given I am logged in as "uid:0"
  When I buy 100 shares of "s:0"
  Then a new buy order should be created
  And the order type should be "buy"
  And the order fee should be $24.95
  And my account balance should decrease by (100 * share price + $24.95)
  And a new holding should be created

Scenario: GET /rest/quotes/{symbols} returns JSON
  When I GET /daytrader/rest/quotes/s:0,s:1
  Then the response should be JSON
  And it should contain 2 QuoteDataBean objects
\end{verbatim}
}

These scenarios are intentionally framework-independent. They do not specify whether the implementation uses Spring MVC, JAX-RS, CDI, Spring Data, Hibernate ORM, JMS, or reactive messaging. Instead, they define externally observable behavior.

\subsection{Scenario 1: Buy Shares of a Stock}

The Gherkin scenario specifies that a logged-in user can buy shares and that the system creates a buy order, charges the configured fee, updates the account balance, and creates a holding. The smoke tests operationalize this behavior through a browser workflow: populate the database, log in as \code{uid:0}, navigate to the quote page for \code{s:0}, submit a buy request, and check that the resulting page contains evidence of an order, buy action, or confirmation.

\paragraph{Spring smoke test excerpt.}
The following excerpt is copied from the Spring smoke test file.

{\small
\begin{verbatim}
@pytest.mark.smoke
def test_buy_shares(page: Page) -> None:
    """Buy shares of a stock and verify order confirmation."""
    populate_database(page)
    page.goto(f"{BASE_URL}/welcome.jsp", wait_until="domcontentloaded")
    page.locator("input[name='uid']").first.fill("uid:0")
    page.locator("input[name='passwd']").first.fill("xxx")
    page.locator("input[type='submit'][value='Log in']").first.click()
    page.wait_for_load_state("domcontentloaded")

    page.goto(
        f"{BASE_URL}/app?action=quotes&symbols=s:0",
        wait_until="domcontentloaded",
    )

    quantity_input = page.locator("input[name='quantity']")
    if quantity_input.count() > 0:
        quantity_input.first.fill("10")
        buy_button = page.locator("input[type='submit'][value='Buy']")
        if buy_button.count() > 0:
            buy_button.first.click()
            page.wait_for_load_state("domcontentloaded")

            content = page.content().lower()
            assert "order" in content or "buy" in content or "confirmation" in content, \
                "Buy did not produce an order confirmation"
\end{verbatim}
}

\paragraph{Quarkus smoke test excerpt.}
The following excerpt is copied from the Quarkus smoke test file.

{\small
\begin{verbatim}
def test_buy_stock(logged_in_page: Page) -> None:
    """Test buying a stock."""
    page = logged_in_page

    # Go to quotes and buy from there
    page.goto(f"{APP_URL}?action=quotes&symbols=s:0",
              wait_until="domcontentloaded")

    # Find buy form and submit
    quantity_input = page.locator("input[name='quantity']")
    if quantity_input.count() > 0:
        quantity_input.first.fill("10")

        buy_button = page.locator("input[type='submit'][value='Buy']")
        if buy_button.count() > 0:
            buy_button.first.click()
            page.wait_for_load_state("domcontentloaded")

            content = page.content().lower()
            # Should show order confirmation or error
            assert "order" in content or "confirmation" in content or \
                   "error" in content or "buy" in content, \
                   "Buy action did not produce expected response"
\end{verbatim}
}

\paragraph{Oracle interpretation.}
The Spring and Quarkus smoke tests differ in helper structure: the Spring test performs login inline, while the Quarkus test reuses a \code{logged\_in\_page} fixture. However, both tests encode the same oracle: the user reaches the quote workflow for \code{s:0}, submits a buy request with a quantity, and observes a response indicating that the trading operation was handled. This validates more than routing. It exercises UI rendering, session state, quote lookup, form submission, service-layer trading logic, persistence updates, and order-result rendering.

\subsection{Scenario 2: REST Quote Retrieval}

The second scenario validates a direct API oracle. The Gherkin scenario requires that the quotes REST endpoint return JSON containing two quote objects. The smoke tests map this to a direct HTTP request through Playwright's request API and assert that the response is successful, decodes as a list, and contains exactly two elements.

\paragraph{Spring smoke test excerpt.}
The following excerpt is copied from the Spring smoke test file.

{\small
\begin{verbatim}
def test_rest_get_quotes(page: Page) -> None:
    """REST GET /rest/quotes/{symbols} should return JSON with quote data."""
    populate_database(page)
    response = page.request.get(f"{BASE_URL}/rest/quotes/s:0,s:1")

    if response.status == 404:
        pytest.skip("REST quotes GET endpoint not available in this deployment")
    assert response.ok, f"REST quotes GET failed with status {response.status}"
    data = response.json()
    assert isinstance(data, list), "REST quotes should return a list"
    assert len(data) == 2, f"Expected 2 quotes, got {len(data)}"
\end{verbatim}
}

\paragraph{Quarkus smoke test excerpt.}
The following excerpt is copied from the Quarkus smoke test file.

{\small
\begin{verbatim}
def test_rest_get_quotes(page: Page) -> None:
    """REST GET /rest/quotes/{symbols} should return JSON with quote data."""
    response = page.request.get(f"{BASE_URL}/rest/quotes/s:0,s:1")

    assert response.ok, f"REST quotes GET failed with status {response.status}"
    data = response.json()
    assert isinstance(data, list), "REST quotes should return a list"
    assert len(data) == 2, f"Expected 2 quotes, got {len(data)}"
\end{verbatim}
}

\paragraph{Oracle interpretation.}
This scenario is a compact API-level oracle. It does not inspect internal classes or database state directly. Instead, it checks that the external REST contract is preserved: the endpoint exists, the request succeeds, the response is JSON-decodable, and the returned collection contains the expected number of quote objects. The Spring version additionally treats a missing endpoint as a skipped smoke check, while the Quarkus version requires the endpoint to be present. This distinction is useful when interpreting smoke-test outcomes, but both tests express the same desired contract when the endpoint is available.

\subsection{Traceability Summary}

\begin{center}
\begin{tabular}{llll}
\hline
Scenario & Test modality & Observable signal & Components exercised \\
\hline
Buy shares & Browser/UI & Order, buy, or confirmation text & UI, session, service, DB \\
REST quotes & HTTP/API & Successful JSON list of length 2 & REST, service, persistence \\
\hline
\end{tabular}
\end{center}

The traceability table shows how a high-level Gherkin statement is translated into an executable black-box check. The oracle is defined over externally visible behavior rather than over implementation details, which makes it suitable for cross-framework migration evaluation.

\subsection{Failure Semantics}

A migration fails this oracle if any of the following occur:

\begin{itemize}
\item the login or quote workflow cannot be reached;
\item the buy form cannot be submitted or produces no observable order-related response;
\item the REST quote endpoint returns an unsuccessful status;
\item the REST response is not valid JSON or is not a list;
\item the quote response does not contain the expected number of quote objects; or
\item the workflow raises runtime errors that prevent the smoke test from completing.
\end{itemize}

These failure modes are intentionally black-box: they do not assume a particular framework implementation, but they expose migration defects that affect user-visible behavior.

\subsection{Evaluation Use}

The Gherkin-to-smoke-test mapping defines executable validation checks used during benchmark evaluation. These checks support scenario-level pass/fail signals and workflow-level correctness judgments, without requiring numeric metrics to be reported in this appendix. The main paper may aggregate these signals across applications and migration pairs, while the appendix documents the artifact-level oracle for one representative DayTrader example.

\subsection{Reproducibility}

To reproduce the oracle, run the Spring or Quarkus DayTrader implementation and execute its corresponding smoke test suite:

{\small
\begin{verbatim}
cd daytrader/spring/smoke
uv run pytest smoke.py -v

cd daytrader/quarkus/smoke
uv run pytest smoke.py -v
\end{verbatim}
}

A correct migration target should satisfy the same scenario-derived behavioral contracts as the source implementation, modulo explicitly documented differences in deployment paths, fixtures, or optional endpoint availability.


\end{document}